\documentclass[11pt,a4paper]{article}
\usepackage{amsmath,amssymb,bm,bbm}
\usepackage[utf8]{inputenc}
\usepackage{graphicx}
\usepackage{color}
\usepackage[dvipsnames]{xcolor}
\usepackage{cite}
\numberwithin{equation}{section}
\allowdisplaybreaks[3]
\usepackage{braket}
\usepackage{subcaption}
\usepackage[linktocpage=true]{hyperref}
\hypersetup{
    colorlinks=true,
    linkcolor=blue,
    citecolor=blue,
    pdfpagemode=FullScreen,
    }

\edef\restoreparindent{\parindent=\the\parindent\relax}
\usepackage{parskip}
\restoreparindent

\usepackage{tikz}
\usetikzlibrary{arrows,arrows.meta,intersections, calc,positioning,decorations.pathreplacing,decorations.pathmorphing,shapes}
\usetikzlibrary{patterns}
\usetikzlibrary{decorations.markings}
\usetikzlibrary{knots}

\def\m{{\mu}}

\def\ep{{\epsilon}}
\def\d{{\rm d}}

\def\frac#1#2{{#1\over #2}}

\def\s{\sqrt}
\def\i{{\rm i}}

\def\CM{{\cal M}}

\def\be{\begin{equation}}
\def\ee{\end{equation}}
\def\ba{\begin{eqnarray}}
\def\ea{\end{eqnarray}}
\def\de{\partial}

\def\f {\frac}
\def\ti{\tilde}

\def\la{\langle}
\def\lb{\rangle}
\def\ep{\epsilon}

\def\SL{\mathrm{SL}}
\def\SU{\mathrm{SU}}

\newcommand{\tr}{\mathrm{tr}}
\newcommand{\Tr}{\mathrm{Tr}}
\newcommand{\C}{\mathbb{C}}
\def\BB{\mathbb{B}}
\def\BS{\mathbb{S}}

\newcommand{\Smat}[2]{{\mathcal{S}_{#1}}^{#2}}

\begin{document}
\begin{titlepage}

\renewcommand{\thefootnote}{\fnsymbol{footnote}}
\begin{flushright}
\begin{tabular}{l}
YITP-22-20\\
IPMU22-0006\\
\end{tabular}
\end{flushright}

\vfill
\begin{center}

\noindent{\Large \textbf{CFT duals of three-dimensional de Sitter gravity}}

\vspace{1.5cm}

\noindent{Yasuaki Hikida,$^a$ Tatsuma Nishioka,$^b$ Tadashi Takayanagi$^{a,c,d}$ and Yusuke Taki$^a$}

\bigskip

\vskip .6 truecm

\centerline{\it $^a$Center for Gravitational Physics and Quantum Information, Yukawa Institute for Theoretical Physics, }
\centerline{\it Kyoto University,  Kitashirakawa Oiwakecho, Sakyo-ku, Kyoto 606-8502, Japan}

\medskip

\centerline{\it $^b$Department of Physics, Osaka University,
Machikaneyama-Cho 1-1, Toyonaka 560-0043, Japan}

\medskip

\centerline{\it $^c$Inamori Research Institute for Science, 620 Suiginya-cho, Shimogyo-ku, Kyoto 600-8411, Japan}

\medskip

\centerline{\it $^d$Kavli Institute for the Physics and Mathematics of the Universe, University of Tokyo,}
\centerline{\it 5-1-5 Kashiwanoha, Kashiwa, Chiba 277-8582, Japan}

\end{center}

\vfill
\vskip 0.5 truecm
\begin{abstract}
We present a class of dS/CFT correspondence between two-dimensional CFTs and three-dimensional de Sitter spaces. We argue that such a CFT includes an $\SU(2)$ WZW model in the critical level limit $k\to -2$, which corresponds to the classical gravity limit. We can generalize this dS/CFT by considering the $\SU(N)$ WZW model in the critical level limit $k\to -N$, dual to the higher-spin gravity on a three-dimensional de Sitter space. We confirm that under this proposed duality the classical partition function in the gravity side can be reproduced from CFT calculations. We also point out a duality relation known in higher-spin holography provides further evidence. Moreover, we analyze two-point functions and entanglement entropy in our dS/CFT correspondence. Possible spectrum and quantum corrections in the gravity theory are discussed.
\end{abstract}
\vfill
\vskip 0.5 truecm

\setcounter{footnote}{0}
\renewcommand{\thefootnote}{\arabic{footnote}}
\end{titlepage}

\newpage

\hrule
\tableofcontents

\bigskip
\hrule
\bigskip

\section{Introduction}

The most important purpose to study string theory is to formulate a theory of quantum gravity. Indeed, the traditional world-sheet approach in string theory enables us to calculate S-matrices in a flat space with stringy quantum corrections. However, it is much harder to describe string theory on curved spacetimes as the world-sheet theory is not tractable in general. Moreover, such a world-sheet approach is limited to a weak coupling limit of quantum gravity. The modern approach based on holography \cite{tHooft:1993dmi,Susskind:1994vu} provides remarkable progresses on this problem. This allows us to investigate anti-de Sitter (AdS) spaces, the most famous example of curved spacetimes with negative curvature,
even for strongly coupled quantum gravity, based on the AdS/CFT correspondence \cite{Maldacena:1997re,Gubser:1998bc,Witten:1998qj}.

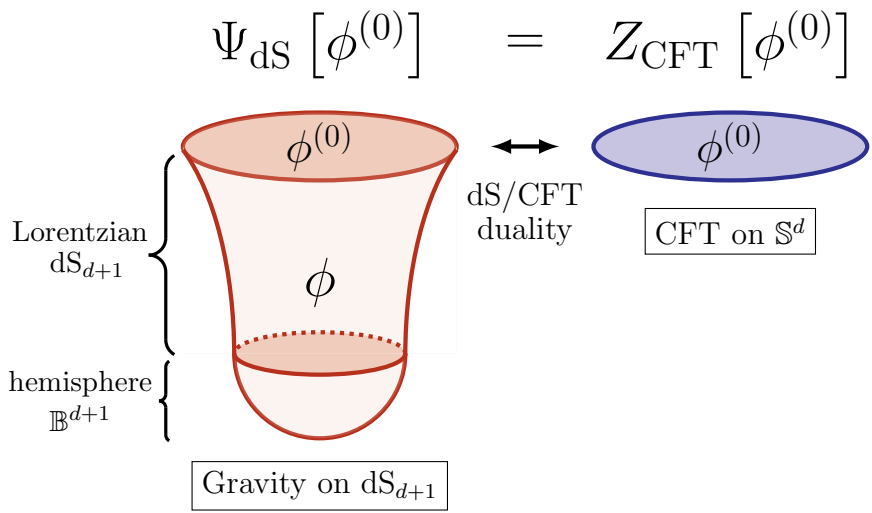
\begin{figure}
  \centering
  \hspace{-2cm}
  \begin{tikzpicture}[>=latex]
    \begin{scope}[scale=0.45]
      \begin{scope}[ultra thick] 
        \draw[BrickRed,fill=BrickRed!20] (0,0) ellipse (4 and 1);
        \draw[draw=none,fill=BrickRed!20,opacity=0.2] (-4,0)--(-4,-6.1)--(4,-6.1)--(4,0)--cycle;
        \draw[BrickRed] (-2.5,-6.1) arc (180:360:2.5);
        \draw[draw=none,fill=BrickRed!20,opacity=0.2] (-2.5,-6.1) arc (180:360:2.5);
        \draw[BrickRed,fill=BrickRed!20] (-2.5,-6.1) arc (180:360:2.5 and 0.625);
        \draw[dotted,BrickRed,fill=BrickRed!20] (2.5,-6.1) arc (0:180:2.5 and 0.625);
        \draw[draw=none,fill=white] (-4,0)--(-4,-6.1)--(-2.5,-6.1)--cycle;
        \draw[draw=none,fill=white] (4,0)--(4,-6.1)--(2.5,-6.1)--cycle;
        \draw[yshift=-0.1cm,BrickRed,fill=white] (-4,0) arc (60:0:3 and 7);
        \draw[yshift=-0.1cm,BrickRed,fill=white] (4,0) arc (120:180:3 and 7);
        \draw[draw=none,fill=BrickRed!20,opacity=0.2] (4,0) arc (0:180:4 and 1);
      \end{scope}
      \node at (0,0) {\LARGE $\phi^{(0)}$};
      \node at (0,-4) {\huge $\phi$};
      \node at (0,3) {\huge $\Psi_{\text{dS}}\left[\phi^{(0)}\right]$};
      \node[draw] at (0,-10) {\large Gravity on dS$_{d+1}$};
      \begin{scope}[xshift=12cm] 
        \draw[ultra thick,Blue,fill=Blue!20] (0,0) ellipse (4 and 1);
        \node[draw] at (0,-2.5) {\large CFT on $\BS^d$};
        \node at (0,3) {\huge $Z_{\text{CFT}}\left[\phi^{(0)}\right]$};
        \node at (0,0) {\LARGE $\phi^{(0)}$};
      \end{scope}
      \node at (6,3) {\huge $=$};
      \begin{scope}[shift={(-4.5,-3.2)},very thick,scale=0.5]
          \draw (-1,0) arc (270:360:1);
          \draw (0,1)--(0,5);
          \draw (0,5) arc (180:120:1);
          \draw (-1,0) arc (90:0:1);
          \draw (0,-1) -- (0,-5);
          \draw (0,-5) arc (180:240:1);
      \end{scope}
      \begin{scope}[shift={(-4.5,-7.5)},very thick,scale=0.2]
          \draw (-1,0) arc (270:360:1);
          \draw (0,1)--(0,5);
          \draw (0,5) arc (180:120:1);
          \draw (-1,0) arc (90:0:1);
          \draw (0,-1) -- (0,-5);
          \draw (0,-5) arc (180:240:1);
      \end{scope}
      \node at (-7,-2.5) {Lorentzian};
      \node at (-6.8,-3.5) {dS$_{d+1}$};
      \node at (-7,-7) {hemisphere};
      \node at (-7,-8) {$\BB^{d+1}$};
      \draw[ultra thick,<->] (5,0) -- (7,0);
      \node at (6,-1.5) {\large dS/CFT};
      \node at (6,-2.5) {\large duality};
    \end{scope}
  \end{tikzpicture}
\caption{A sketch of the dS/CFT correspondence. $\phi$ denotes all bulk fields and $\phi^{(0)}$ does their values at the asymptotic boundary (i.e. future infinity).}
\label{dSCFTfig}
\end{figure}

To understand the real cosmology of the early universe, it is desirable to understand quantum gravity on de Sitter (dS) spaces, the representative of positive curvature spacetimes. This motivates us to study the holographic duality for de Sitter spaces. 
A possible holography for de Sitter spaces has been known as the dS/CFT correspondence 
\cite{Strominger:2001pn,Witten:2001kn,Maldacena:2002vr}. The dS/CFT argues that $(d+1)$-dimensional gravity on a de Sitter space is dual to a $d$-dimensional Euclidean conformal field theory (CFT) living on its space-like boundary (i.e. the future/past infinity), which may be deduced from the matching of the geometrical symmetry of de Sitter spaces and that of conformal symmetry. Consider the Hartle-Hawking wave functional of gravity on a $(d+1)$-dimensional de Sitter space, starting from the Euclidean tunneling region of a hemisphere (radius $L$):
\ba
\d s^2=L^2\left(\d\tau^2+\cos\tau^2\, \d\Omega_d^2\right) \, , \qquad
\left(-\frac{\pi}{2}\leq \tau\leq 0 \right) \ ,
\ea
and continuing to the Lorentzian de Sitter spacetime (radius $L$)
\ba
\d s^2=L^2\left(-\d T^2+\cosh^2 T\, \d\Omega_d^2\right) \, , \qquad (T\geq 0) \ ,
\ea
by setting $\tau=\i \, T$, where $\d\Omega_d^2$ is the metric of $d$-dimensional sphere with a unit radius.  The dS/CFT equates this Hartle-Hawking wave functional $\Psi_\text{dS}$ with the partition function of the dual CFT $Z_\text{CFT}$ on $\mathbb{S}^d$ \cite{Maldacena:2002vr}, as depicted in figure \ref{dSCFTfig}. 

In the dS/CFT, we expect that the dual Euclidean CFT is {\it exotic} in some sense because the standard Euclidean holographic CFT should be dual to gravity on a Euclidean AdS or equally hyperbolic space. For example, the dS/CFT predicts that dual $d$-dimensional CFTs should have imaginary valued central charges when $d$ is even \cite{Maldacena:2002vr}. An analogous result has been found for the 
holographic entanglement entropy \cite{Ryu:2006bv} in the de Sitter spacetime \cite{Narayan:2015vda,Sato:2015tta,Miyaji:2015yva}, which is due  to the fact that there is no space-like geodesic which connects a pair of points on the space-like boundary. Another way to mention the difference between the AdS/CFT and the dS/CFT is that the Wick rotation of a dS into its Euclidean space (i.e., a sphere) largely changes the asymptotic boundary structure as there is no boundary on a sphere, while the Wick rotation of an AdS to a hyperbolic space does not change its asymptotic boundary. This difference is quite important because the holography relates the bulk gravity to its boundary in general.

As opposed to the AdS/CFT, 
there have been only a few explicit and microscopic examples of dS/CFT. This is partly because the embedding of de Sitter spaces into string theory has been known to be a highly complicated problem and partly because the dual CFT is expected to be unusual as we mentioned.
The only known microscopic example of dS/CFT, as far as we are aware of, was proposed in \cite{Anninos:2011ui}, which states a duality between higher-spin gravity on dS$_4$ and 3d Sp$(N)$ vector model. This duality may be regarded as an ``analytic continuation'' of Klebanov-Polyakov duality \cite{Klebanov:2002ja} between higher-spin gravity on AdS$_4$ and 3d O$(N)$ vector model. Indeed the  Sp$(N)$ vector model is exotic in that it consists of fermionic scalar fields. Since the dual gravity theory includes the infinite tower of higher-spin fields in this example, a CFT dual of Einstein gravity on de Sitter spaces has not been available so far. 

The main aim of this paper is to provide the first example of CFT dual of Einstein gravity on a de Sitter space. We focus on the gravity on a three-dimensional de Sitter space (dS$_3$), where the dual Euclidean CFT lives in two dimensions. In this lower dimensional example, we have the advantage that the de Sitter gravity is described by a three-dimensional Chern-Simons gauge theory \cite{Witten:1988hc}. Moreover, a Chern-Simons gauge theory is also known to be equivalent to a two-dimensional CFT \cite{Witten:1988hf}. Combining these famous facts with a twist,
we will obtain a class of microscopic CFT duals of gravity on dS$_3$. In this lower dimensional setup, the symmetry of infinite-dimensional algebra of 2d CFT helps us to solve the theory exactly.
A part of our results has been already reported in the letter version \cite{Hikida:2021ese}. In this paper, we will give extensive evidences for our new example of dS$_3/$CFT$_2$ from the viewpoints of partition function, two-point functions, entanglement entropy and higher-spin holography, paying much attention to extension of the Euclidean dS$_3$ (hemisphere $\BB^3$) to Lorentzian dS$_3$.

Before we proceed, we would like to mention that there have been other approaches to study holography in de Sitter spaces. They include
the recent progress in the light of the dS/dS correspondence \cite{Alishahiha:2004md,Dong:2018cuv,Gorbenko:2018oov,Geng:2021wcq},
the use of the surface/state duality \cite{Miyaji:2015yva}, and possible holographic duality for the dS static patches \cite{Susskind:2021dfc,Susskind:2021esx}. In all these examples, the dual non-gravitational theories are expected to be less exotic, such as the $T\bar{T}$ deformations of CFTs. One more interesting approach is the holographic duality proposed in \cite{Freivogel:2006xu}, where the dual CFT is localized on a codimension-two space. Even though these other approaches look different from the original dS/CFT \cite{Strominger:2001pn,Witten:2001kn,Maldacena:2002vr} in that their CFT dual lives on time-like surfaces, it is still possible that future progress may connect them, which is beyond the scope of this paper. In this paper, we study CFT duals of dS$_3$, which fit nicely with the original dS/CFT proposal.

\subsection{Our proposal of \texorpdfstring{$\text{dS}_3/\text{CFT}_2$}{dS3/CFT2} correspondence}
\label{sec:dSCFT}

Here we would like to summarize our new holographic proposal from the beginning for readers' convenience. Our proposal of $\text{dS}_3/\text{CFT}_2$ correspondence can be summarized as follows in general:
\ba\label{dualitytwo}
    \parbox{5cm}{\centering Einstein gravity on dS$_3$\\ \medskip with $\displaystyle\frac{L}{G_N} \gg1$}      
    \qquad \Longleftrightarrow \qquad
    \parbox{6cm}{\centering $\widehat{\SU(2)}_k\times M_\text{CFT}$ \\ \medskip in $\displaystyle k\to -2+\i\,\frac{4G_N}{L} ~ (\simeq -2)$} \, ,
\ea
where $L$ and $G_N$ are the radius of dS$_3$ and the Newton constant, respectively. Moreover, $\widehat{\text{SU}(2)}_k$ describes the SU$(2)$ Wess-Zumino-Witten (WZW) model at level $k$ and $M_\text{CFT}$ is a certain two-dimensional CFT. The product $\widehat{\text{SU}(2)}_k\times M_\text{CFT}$ allows warped ones, and
the choice of $M_\text{CFT}$ is dual to that of the matter fields in three-dimensional gravity. Note that in this $k\to -2$ limit, which is dual to the limit $L/G_N\to \infty$, the dominant contributions to physical quantities such as the free energy is dominated by $\widehat{\SU(2)}_k$ because its central charge gets divergent
\ba
c=\frac{3k}{k+2}\simeq \i\,\frac{3L}{2G_N}~ (\equiv \i\, c^{(g)})\to \i\,\infty \, .
\label{centtwo}
\ea
Notice also that the central charge takes imaginary value in agreement with 
general argument of dS$_3/$CFT$_2$ \cite{Maldacena:2002vr}. Therefore, the contributions from the $M_\text{CFT}$ part can be negligible when we compare the physical quantities with the gravity sector in the classical limit.
A primary state in the CFT with the energy $\Delta=h+\bar{h}$ (i.e.\,the conformal dimension of the corresponding primary operator) 
is dual to the bulk excitation with energy
\ba
\Delta=\i\, E^{(g)}L\ (\equiv \i\, \Delta^{(g)}) \, . \label{energyds}
\ea

We can generalize this into a higher-spin version of dS$_3/$CFT$_2$:
\begin{align}
\parbox{7cm}{\centering Spin $s \, (=2,3,\ldots,N)$ gravity on dS$_3$\\ \medskip with  $\displaystyle \frac{L}{G_N} \gg1$} 
    \quad \Longleftrightarrow\quad
\parbox{7.5cm}{\centering $\widehat{\text{SU}(N)}_k\times M_\text{CFT}$\\ \medskip in  $\displaystyle k\to -N+\i\,\frac{16\, \epsilon_N^2\, G_N}{L}~(\simeq -N)$} \label{dualityN}
\end{align}
with
\begin{align}
\epsilon_N = \frac{1}{12} N (N^2 -1) \, . \label{epsilonN}
\end{align}
In this duality, the central charge of the two-dimensional CFT again behaves as follows:
\ba\label{centNa}
c=\frac{k(N^2-1)}{k+N}\simeq \i\, \frac{3L}{4 \epsilon_N G_N} \ (\equiv \i\, c^{(g)})\to \i \, \infty \, .
\ea
Here, as the spin $s \, (=2,3,\ldots, N)$ gravity  in (\ref{dualityN}), we consider the higher-spin gravity on $\mathbb{S}^3$ described by $\text{SU}(N) \times \text{SU}(N)$ Chern-Simons gauge theory.  The Einstein gravity on $\mathbb{S}^3$ corresponds to the case with $N=2$ \cite{Witten:1988hc}.
To describe a gravity theory on $\mathbb{S}^3$, the Chern-Simons coupling should take an imaginary value as $k_\text{CS} = \i \,  \kappa$. The semiclassical limit corresponds to $\kappa\to \infty$ and for $N=2$ this gives the Einstein gravity limit
(see \cite{Banados:1998tb,Castro:2011xb,Castro:2012gc,Cotler:2019nbi,Castro:2020smu,Anninos:2020hfj,Anninos:2021ihe} for various studies of this limit).

A special example of the proposed duality  (\ref{dualitytwo}) and (\ref{dualityN}) 
with the $M_\text{CFT}$ part specified, can be obtained from a
two-dimensional $W_N$-minimal model described by
\begin{align}
W_{N,k}\equiv\frac{\text{SU}(N)_k\times \text{SU}(N)_1}{\text{SU}(N)_{k+1}}  \label{cosetc}
\end{align}
with the central charge 
\begin{align}
c=(N-1)\left(1-\frac{N(N+1)}{(N+k)(N+k+1)}\right) \, . \label{centwm}
\end{align}
We argue that in the $k\to -N$ limit of (\ref{dualityN}), this (analytically continued) $W_N$-minimal model is dual to a higher-spin gravity on conical defects coupled to a complex scalar field. We can also regard this as an ``analytic continuation'' of Gaberdiel-Gopakumar duality \cite{Gaberdiel:2010pz} between higher-spin gravity on AdS$_3$ and the $W_N$-minimal model.

As emphasized above, the SU$(N)_k$ part of the coset \eqref{cosetc} dominates at the limit $k \to -N$.
On the other hand, the coset with generic $k$ has been believed to be equivalent to Toda field theory, which was recently confirmed in \cite{Creutzig:2021ykz}. The Toda field theory has a parameter $b$ and the large central charge is realized at the limit $b \to 0$ (or $b \to \infty$ via the self-duality). This implies that Toda field theory with $b \to 0$ is equivalent to SU$(N)$ WZW model with $k \to -N$ at the leading order in $1/c^{(g)}$. For the simplest case with $N=2$, Liouville field theory with $b \to 0$ is equivalent to SU(2) WZW model with $k \to -2$ at the leading order in $1/c^{(g)}$. We sometimes use Liouville/Toda description instead of that of SU$(N)$ WZW model since the former is more convenient than the latter for some purposes. See appendix \ref{app:DS} for details.

According to \cite{Maldacena:2002vr}, a formula of dS/CFT correspondence may be expressed as the equality between the Hartle-Hawking wave functional 
$\Psi_\text{dS}$ for dS$_3$ and the partition function of the dual CFT$_2$: 
\begin{align}
\Psi_\text{dS} \left[\phi^{(0)}\right]=Z_\text{CFT} \left[\phi^{(0)}\right] \, , \label{dsMaldacena}
\end{align}
where  $\phi^{(0)}$ symbolically describes the values of all bulk fields $\phi$ restricted to the future boundary of dS$_3$. This is also interpreted as the external source in the dual CFT$_2$ (refer to figure \ref{dSCFTfig}). This is a natural extension of the bulk-boundary relation \cite{Gubser:1998bc,Witten:1998qj} for the AdS/CFT.

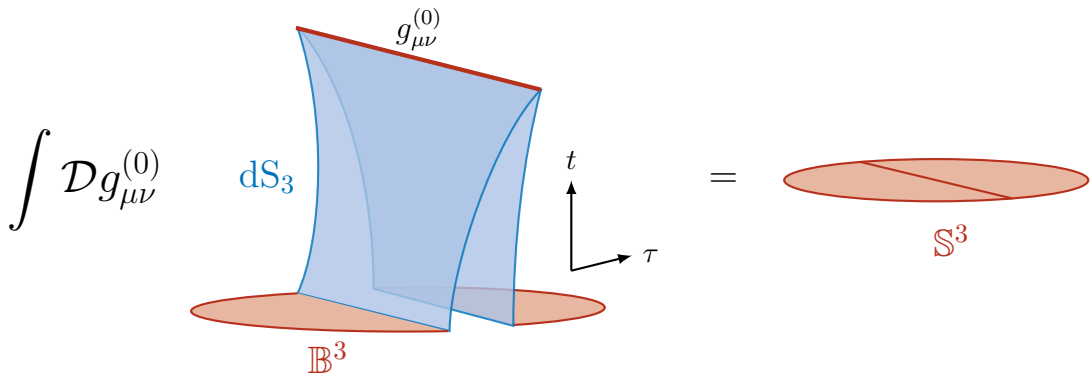
\begin{figure}
  \center
  \begin{tikzpicture}[thick,scale=0.4, >=latex]
    \begin{scope} 
      \draw[BrickRed,fill=BrickRed!30] (3.1,-8.8) arc (-60:105:6 and 0.67);
      \draw[RoyalBlue,fill=RoyalBlue!30,opacity=0.8] (-1.5,-7.6) arc (0:59:5.1 and 10.1)--(4,-1)--(4,-1) arc (146:180:5.3 and 14)--cycle;
      \node[below, BrickRed] at (-3, -9) {\Large $\BB^3$};
      \node[below, RoyalBlue] at (-5, -3) {\Large $\text{dS}_3$};
    \end{scope}
    \begin{scope} 
      \draw[RoyalBlue,fill=RoyalBlue!30,opacity=0.8] (-4,1.05) .. controls (-3,-2) and (-3.2,-6) .. (-4,-7.75)--(1,-9) .. controls (1,-7) and (2.5,-2.5) .. (4,-1) --cycle ;
      \draw[BrickRed,fill=BrickRed!30] (-4,-7.75) arc (120:282:7 and 0.68);
    \end{scope}
    \draw[BrickRed,ultra thick] (-4.05,1.05)--(4,-1);
    \node at (-11,-4) {\LARGE $\displaystyle  \int\mathcal{D}g^{(0)}_{\mu\nu}$};
    \node at (0,1) {\large $\displaystyle g^{(0)}_{\mu\nu}$};
    \begin{scope}[xshift=-5cm, yshift=-2cm] 
      \draw[->] (10,-5)--(10,-2); 
      \draw[->] (10,-5)--(12,-4.5);
      \node[above] at (10,-2) {$t$}; 
      \node[right] at (12,-4.5) {$\tau$};
    \end{scope}
    \begin{scope}[xshift=17cm, yshift=-4cm]
        \node at (-7,0) {\Large $=$};
        \draw[BrickRed,fill=BrickRed!30] (5,0) arc (0:360:5 and 0.7);
        \draw[BrickRed,-] (120:5 and 0.7) -- (-60:5 and 0.7);
        \node[BrickRed] at (0.5,-2) {\Large $\BS^3$};
    \end{scope}
  \end{tikzpicture}
  \caption{A sketch of the path integral in the right-hand side of \eqref{square}. The red and blue regions represent the Euclidean $(\BB^3)$ and Lorentzian $(\text{dS}_3)$ part of Hartle-Hawking wave functional $\Psi_{\text{dS}}$, respectively.  The Lorentzian parts of $\Psi_{\text{dS}}$ and its dual $\Psi_{\text{dS}}^*$ cancel out, leaving only the Euclidean part $Z_{\text{G}}\left[\BS^3\right]$. }
  \label{fig:norm}
\end{figure}

To test the new duality proposed above, we will mainly work 
in a Euclidean version of the dS/CFT, namely the duality between gravity on $\mathbb{S}^3$
and the $k\to -N$ limit of the two-dimensional CFT described in  (\ref{dualitytwo}) and (\ref{dualityN}). This is formally related to the Lorentzian version (\ref{dsMaldacena}) by gluing two copies of de Sitter space or equally taking the inner product of the Hartle-Hawking wave functional (see figure \ref{fig:norm}):
\begin{align}
Z_\text{G}\left[\mathbb{S}^3\right] = \int \mathcal{D} g_{\mu\nu}^{(0)}\, \left| \Psi_\text{dS} \left[g_{\mu\nu}^{(0)}\right]\right|^2 \, , \label{square}
\end{align}
where $g_{\mu\nu}^{(0)}$ is the metric on the future boundary of de Sitter space.
Indeed, the contribution from the Lorentzian dS$_3$ gives only a phase factor to the wave functional:
\begin{align}
    \Psi_\text{dS} \left[g_{\mu\nu}^{(0)}\right]
    \sim
    \exp\left[ \i\,I_\textrm{G}^{(\textrm{L})}\left[\text{dS}_3\right] - I_\textrm{G}^{(\textrm{E})}\left[\BB^3\right] \right] \ ,
\end{align}
where $I_\textrm{G}^{(\textrm{L})}$ and $I_\textrm{G}^{(\textrm{E})}$ are Lorentzian and Euclidean gravity actions, respectively.
Hence, the Lorentzian part cancels out in the square of the wave functional and the Euclidean part leads to the gravity partition function on $\mathbb{S}^3$ as in \eqref{square} (see figure \ref{fig:norm}). 
In the semiclassical limit $L/G_N\gg1$, a saddle point solution dominates in the path integral over the boundary metric $g_{\mu\nu}^{(0)}$ in the right hand side of \eqref{square}.
For the Hartle-Hawking wave functional, the saddle solution for the boundary metric is $\BS^2$.
Then it follows from the dS/CFT dictionary \eqref{dsMaldacena} that the gravity partition function is given by the square of the dual CFT partition function in the semiclassical limit:
\begin{align}\label{partitionappro}
    Z_{\text{G}}\left[\BS^3\right]\simeq \left|\Psi_{\text{dS}}\left[\BS^2\right]\right|^2 = \left|Z_{\text{CFT}}\left[\BS^2\right]\right|^2\, .
\end{align}

In our proposal, the CFT on the right-hand side of \eqref{partitionappro} is the non-chiral $\widehat{\text{SU}(2)}_k$ WZW model. 
Both the holomorphic and anti-holomorphic parts of the WZW model are equivalent to $\text{SU(2)}$ Chern-Simons theories of level $k$ and $-k$ respectively on a manifold with boundary $\BS^2$ \cite{Witten:1988hf,Coussaert:1995zp,Donnay:2016iyk}:
\begin{align}\label{CS_WZW}
    Z_{\SU(2)_k\, \text{WZW}}\left[\BS^2\right]
    =
    Z_{\SU(2)_k\times \SU(2)_{-k}\, \text{CS}}\left[\BB^3\right]\, .
\end{align}
Squaring both sides and gluing two copies of $\BB^3$ along the boundaries $\BS^2$ for the Chern-Simons theory, \eqref{CS_WZW} yields
\begin{align}\label{CS_WZWA}
    \left| Z_{\SU(2)_k\, \text{WZW}}\left[\BS^2\right]\right|^2
    =
    Z_{\SU(2)_k\times \SU(2)_{-k}\, \text{CS}}\left[\BS^3\right]
    =
    \left|Z_{\SU(2)_k\, \text{CS}}\left[\BS^3\right]\right|^2
    \ . 
\end{align}
Substituting into \eqref{partitionappro}, we find the relation:
\begin{align}\label{partzero}
    Z_{\text{G}}\left[\BS^3\right]\simeq\left|Z_{\SU(2)_k\,\text{CS}}\left[\BS^3\right]\right|^2=\left|\Smat{0}{0}\right|^2\, ,
\end{align}
where $\Smat{0}{0}$ is the identity component of the modular $\mathcal{S}$-matrix of the character of the SU$(N)_k$ WZW model.
This relation provides the simplest consequence of our dS/CFT proposal.
As we will see in this paper, we can generalize (\ref{partzero}) to excited states by taking into account Wilson lines.

In this way, the proposed dS/CFT duality allows us to calculate various physical quantities in the semiclassical gravity on $\mathbb{S}^3$ from those in the WZW model in the diverging central charge limit $c\to \i \, \infty$. 
To promote this Euclidean setup to the dS/CFT for the Lorentzian de Sitter space, we need to perform a Wick rotation. 
We will discuss this issue in the context of calculations of partition function, two-point function, and entanglement entropy later. 
Notice that this new duality is highly nontrivial in that it cannot be obtained by simply combining the well-known results of \cite{Witten:1988hc} and  \cite{Witten:1988hf}.
While the relation \eqref{partzero} takes exactly the same form, the semiclassical limit $|k_{\text{CS}}|=\kappa\to \infty$ of the Chern-Simons formulation of Einstein gravity only leads to a finite central charge $c\simeq N^2-1$ and does not fit with the general expectation of the dS/CFT that we should have $c\to \i \, \infty$.

\subsection{Organization of this paper}
The paper is organized as follows. In the next section, we start with the detailed statement of 
our proposal of dS/CFT for three-dimensional Einstein gravity, summarized as
(\ref{dualitytwo}). We will calculate the gravity partition functions in the presence of various excitations in both the CFT and gravity sides. Then we confirm that both calculations match perfectly in the Einstein gravity limit. In passing, we will comment on the relation between the topological entanglement entropy and the de Sitter entropy.

In section \ref{sec:hs}, we present a higher-spin generalization of the three-dimensional dS/CFT, summarized as (\ref{dualityN}). We explicitly evaluate the partition functions in three-dimensional higher-spin gravity for various classical solutions based on two different approaches and show that they perfectly agree with the CFT results computed from the $\mathcal{S}$-matrices.

In section \ref{sec:2pt}, we study the effects of Lorentzian dS$_3$ on our dS/CFT duality by examining partition functions, two-point functions, and entanglement entropies. We first compute the classical partition function of Liouville CFT, which is found to be consistent with previous result.
We then show that an appropriate analytical continuation gives two-point functions in the dual CFTs if we take into account an exotic UV cutoff prescription. This shows a way how we extend our original Euclidean proposal to the Lorentzian dS/CFT. We will also give the holographic entanglement entropy in our dS/CFT and confirm our prediction in the light of brane-world holography.

In subsection \ref{sec:GG}, we consider a special version of dS/CFT given in terms of the $W_N$-minimal model CFT. We interpret our dS/CFT as an analytical continuation of Gaberdiel-Gopakumar duality. We also show that the duality relation known as the triality in the $W_N$-minimal model gives us further evidence of our dS/CFT.

In section \ref{sec:spectrum}, we discuss the possible spectrum of WZW model and its geometrical interpretation.
In section \ref{sec:qc}, we work out CFT predictions about quantum corrections in the gravity theory on dS$_3$ at the one-loop order.
In section \ref{sec:conclusion}, we summarize our conclusions and discuss future problems.
In appendix \ref{app:DS}, we give an interpretation of our results in terms of the Liouville/Toda CFT. In appendix \ref{app:Wilsonline}, we presents calculations of the expectation values of Wilson loops and their connection to geodesic distances.

\section{Einstein gravity on \texorpdfstring{$\mathbb{S}^3$}{S3} and \texorpdfstring{$\text{dS}_3/\text{CFT}_2$}{dS3/CFT2}}
\label{sec:gravity}

In this and next sections, we compute the gravity partition functions \eqref{square} both from the large central charge limit of the WZW model and from the classical (higher-spin) gravity. We then compare them and find perfect matches. In this section, we start from the simplest case with $N=2$, where the CFT is given by the SU$(2)$ WZW model and the classical gravity is defined by the Einstein-Hilbert action. The goal in this section is to confirm the proposed duality relation (\ref{dualitytwo}).

\subsection{CFT description}

Consider the SU$(2)$ WZW model at level $k$ and the Hilbert space created from the $(2j+1)$-dimensional representation denoted by $R_j$. 
Its character in the representation $R_j$ is defined as
\begin{align}
\chi_j (\tau) = \text{Tr}_{R_j} \left[e^{2 \pi \i \tau \left(L_0-\frac{c}{24}\right)} \right]\, , 
\end{align}
where $L_0$ is the zero mode of energy momentum tensor. Under the modular transformation $\tau \to - 1/\tau$, the characters transform as
\begin{align}
\chi_j (-1/\tau) = \sum_{l}\, \Smat{j}{l}\, \chi_l (\tau) \, , 
\end{align}
where the $\mathcal{S}$-matrix for the SU$(2)$ WZW model is given by%
\footnote{The modular $\mathcal{S}$-matrix is computed for integer $k$ and performed an analytic continuation for complex $k$. This procedure usually works well as in \cite{Anninos:2020hfj} but may require more justification. The same is true for higher rank expression in \eqref{SN} below. Supporting arguments are given in appendix \ref{app:DS} by utilizing Liouville/Toda description as discussed in the introduction.}
\begin{align}
\Smat{j}{l}=\s{\frac{2}{k+2}}\,\sin\left[\frac{\pi}{k+2}\,(2j+1)\,(2l+1)\right] \, .  \label{SU2S}
\end{align}

\begin{figure}
  \centering
    \begin{tikzpicture}[scale=0.7, every node/.style={scale=0.8}, xshift=-3cm]
        \begin{scope}[thick]
            \begin{scope}[xshift=-1cm]
             \begin{scope}[scale=0.4]
             \draw[fill=lightgray!20!white] (0,0) ellipse (3 and 1.5);
                \begin{scope}
                \clip (0,-1.8) ellipse (3 and 2.5);
                \draw (0,2.2) ellipse (3 and 2.5);
                \end{scope}
                \begin{scope}
                \clip (0,2.2) ellipse (3 and 2.5);
                \draw[fill=white] (0,-2.2) ellipse (3 and 2.5);
                \end{scope}
                \draw[very thick, RoyalBlue] (0, 0) ellipse (2.3 and 1);
            \end{scope}
            \begin{scope}[scale=0.4, xshift=8cm, rotate=90]
                \draw[fill=lightgray!20!white] (0,0) ellipse (3 and 1.5);
                \begin{scope}
                \clip (0,-1.8) ellipse (3 and 2.5);
                \draw (0,2.2) ellipse (3 and 2.5);
                \end{scope}
                \begin{scope}
                \clip (0,2.2) ellipse (3 and 2.5);
                \draw[fill=white] (0,-2.2) ellipse (3 and 2.5);
                \end{scope}
                \draw[very thick, RoyalBlue] (0, 0) ellipse (2.3 and 1);
            \end{scope}
            \node at (-2.2, -0.1) {\Large $\displaystyle \sum_i \Smat{i}{j}$};
            \node[RoyalBlue] at (0, 1) {\Large $R_i$};
            \node[RoyalBlue] at (2.3, -0.8) {\Large $R_j$};
            \node at (1.8, 0) {\Large $=$};
            \node at (0, -1) {\large $\tau$};
            \node at (4.3, -1) {\large $\displaystyle -\frac{1}{\tau}$};
        \end{scope}
        \begin{scope}[yshift=-3.5cm]
             \begin{scope}[scale=0.4]
                \draw[fill=lightgray!20!white] (0,0) ellipse (3 and 1.5);
                \begin{scope}
                \clip (0,-1.8) ellipse (3 and 2.5);
                \draw (0,2.2) ellipse (3 and 2.5);
                \end{scope}
                \begin{scope}
                \clip (0,2.2) ellipse (3 and 2.5);
                \draw[fill=white] (0,-2.2) ellipse (3 and 2.5);
                \end{scope}
                \draw[very thick, RedOrange] (0, 0) ellipse (2.3 and 1);
            \end{scope}
            \begin{scope}[scale=0.4, xshift=7.3cm, rotate=90]
                \draw[fill=lightgray!20!white] (0,0) ellipse (3 and 1.5);
                \begin{scope}
                \clip (0,-1.8) ellipse (3 and 2.5);
                \draw (0,2.2) ellipse (3 and 2.5);
                \end{scope}
                \begin{scope}
                \clip (0,2.2) ellipse (3 and 2.5);
                \draw[fill=white] (0,-2.2) ellipse (3 and 2.5);
                \end{scope}
            \end{scope}
            \draw (-3.5, 0) circle (1);
            \draw[dotted] (-3.5, 0) ellipse (1 and 0.3);
            \draw[very thick, RedOrange] (-3.5, 0.2) ellipse (0.7 and 0.35);
            \node[RedOrange] at (-3.5, -0.5) {\Large $R_i$};
            \node at (-1.9, 0) {\Large $=$};
            \node[RedOrange] at (0, 1) {\Large $R_i$};
            \node at (1.7, 0) {\Large $\cdot$};
            \node at (4.5, 0) {\Large $=~ \Smat{0}{i}$};
        \end{scope}
        \begin{scope}[yshift=-7cm]
             \begin{scope}[scale=0.4]
                \draw[fill=lightgray!20!white] (0,0) ellipse (3 and 1.5);
                \begin{scope}
                \clip (0,-1.8) ellipse (3 and 2.5);
                \draw (0,2.2) ellipse (3 and 2.5);
                \end{scope}
                \begin{scope}
                \clip (0,2.2) ellipse (3 and 2.5);
                \draw[fill=white] (0,-2.2) ellipse (3 and 2.5);
                \end{scope}
                \draw[very thick, RedOrange] (0, 0) ellipse (2.3 and 1);
            \end{scope}
            \begin{scope}[scale=0.4, xshift=7.3cm, rotate=90]
                \draw[fill=lightgray!20!white] (0,0) ellipse (3 and 1.5);
                \begin{scope}
                \clip (0,-1.8) ellipse (3 and 2.5);
                \draw (0,2.2) ellipse (3 and 2.5);
                \end{scope}
                \begin{scope}
                \clip (0,2.2) ellipse (3 and 2.5);
                \draw[fill=white] (0,-2.2) ellipse (3 and 2.5);
                \end{scope}
                \draw[very thick, RoyalBlue] (0, 0) ellipse (2.3 and 1);
            \end{scope}
            \draw (-3.5, 0) circle (1);
            \draw[dotted] (-3.5, 0) ellipse (1 and 0.3);
            \begin{knot}[flip crossing=1]
                \strand[very thick, RedOrange] (-3.7, 0) ellipse (0.6 and 0.3);
                \strand[very thick, RoyalBlue] (-3, 0) ellipse (0.3 and 0.6);
            \end{knot}
            \node[RedOrange] at (-3.7, -0.6) {\Large $R_i$};
            \node[RoyalBlue] at (-3.6, 0.6) {\Large $R_j$};
            \node at (-1.9, 0) {\Large $=$};
            \node[RedOrange] at (0, 1) {\Large $R_i$};
            \node[RoyalBlue] at (2, -0.8) {\Large $R_j$};
            \node at (1.7, 0) {\Large $\cdot$};
            \node at (4.5, 0) {\Large $=~ \Smat{i}{j}$};
        \end{scope}
    \end{scope}
    
    \begin{scope}[xshift=11cm, yshift=-1cm, thick]
        \begin{scope}[xshift=0.1cm]
            \begin{scope}[xshift=0.5cm]
                \draw (-3, -0.25) ellipse (1 and 0.5);
                \draw (-3, 0.25) arc (90:270:0.1 and 0.5);
                \draw[draw=none, fill=lightgray!20!white] (-3, -0.25) ellipse (0.1 and 0.5);
                \draw[dotted] (-3, -0.75) arc (-90:90:0.1 and 0.5);
                \node at (-3.5, -0.25) {\large $\CM'$};
                \node at (-2.5, -0.25) {\large $\CM''$};
            \end{scope}
            
            \node at (-0.8, -0.25) {\Large $=$};
            
            \draw (1, 1) arc (90:270:1 and 0.5);
            \draw[draw=none, fill=lightgray!50] (1, 0.5) ellipse (0.1 and 0.5);
            \draw[fill=RoyalBlue, fill opacity=0.5] (1, 1) arc (90:270:0.1 and 0.5);
            \draw[dotted] (1, 0) arc (-90:90:0.1 and 0.5);
            \draw[fill=RoyalBlue, fill opacity=0.5] (1, 0) arc (-90:90:0.5 and 0.5);
            \node at (0.5, 0.5) {\large $\CM'$};
            
            \node at (2, 0.5) {\Large $\times$};
            
            \draw (3, 0) arc (-90:90:1 and 0.5);
            \draw[fill=RoyalBlue, fill opacity=0.5] (3, 1) arc (90:270:0.5 and 0.5);
            \draw[draw=none, fill=lightgray!20!white]
            (3, 0.5) ellipse (0.1 and 0.5);
            \draw (3, 1) arc (90:270:0.1 and 0.5);
            \draw[dotted] (3, 0) arc (-90:90:0.1 and 0.5);
            \node at (3.5, 0.5) {\large $\CM''$};
            
            \draw (0, -0.25) --+ (4, 0);
            
            \draw[fill=RoyalBlue, fill opacity=0.5] (2, -1) circle (0.5);
            \draw (2, -0.5) arc (90:270:0.1 and 0.5);
            \draw[dotted] (2, -1.5) arc (-90:90:0.1 and 0.5);
            \node at (1.1, -1) {\Large $\BS^3$};
        \end{scope}
    
        \begin{scope}[yshift=-4cm]
            \begin{scope}[xshift=0.7cm]
                \draw (-3, -0.25) circle (1);
                \draw[dotted] (-3, -0.25) ellipse (1 and 0.3); 
                \draw[RedOrange, very thick] (-3.5, -0.25) ellipse (0.3 and 0.15);
                \draw[RoyalBlue, very thick] (-2.5, -0.25) ellipse (0.3 and 0.15);
                \node[RedOrange] at (-3.5, -0.8) {\large $R_i$};
                \node[RoyalBlue] at (-2.5, -0.8) {\large $R_j$};
             \end{scope} 
            \node at (-0.7, -0.25) {\Large $=$};
            \draw (1, 0.5) circle (0.5);
            \draw[dotted] (1, 0.5) ellipse (0.5 and 0.15);
            \draw[RedOrange, very thick] (1, 0.65) ellipse (0.3 and 0.15);
            \node[RedOrange] at (1, 0.25) {\large $R_i$};
            \node at (2, 0.5) {\Large $\times$};
            \draw (3, 0.5) circle (0.5);
            \draw[dotted] (3, 0.5) ellipse (0.5 and 0.15);
            \draw[RoyalBlue, very thick] (3, 0.65) ellipse (0.3 and 0.15);
            \node[RoyalBlue] at (3, 0.25) {\large $R_j$};
            \draw (0, -0.25) --+ (4, 0);
            \draw (2, -1) circle (0.5);
            \draw[dotted] (2, -1) ellipse (0.5 and 0.15);
            \node at (5.5, -0.25) {\Large $\displaystyle = ~ \frac{\Smat{0}{i}\, \Smat{0}{j}}{\Smat{0}{0}}$};
        \end{scope}
    \end{scope}
    \end{tikzpicture}
\caption{Sketches of the computations of partition functions 
in Chern-Simons theory on $\mathbb{S}^3$.}
\label{wilsa}
\end{figure}
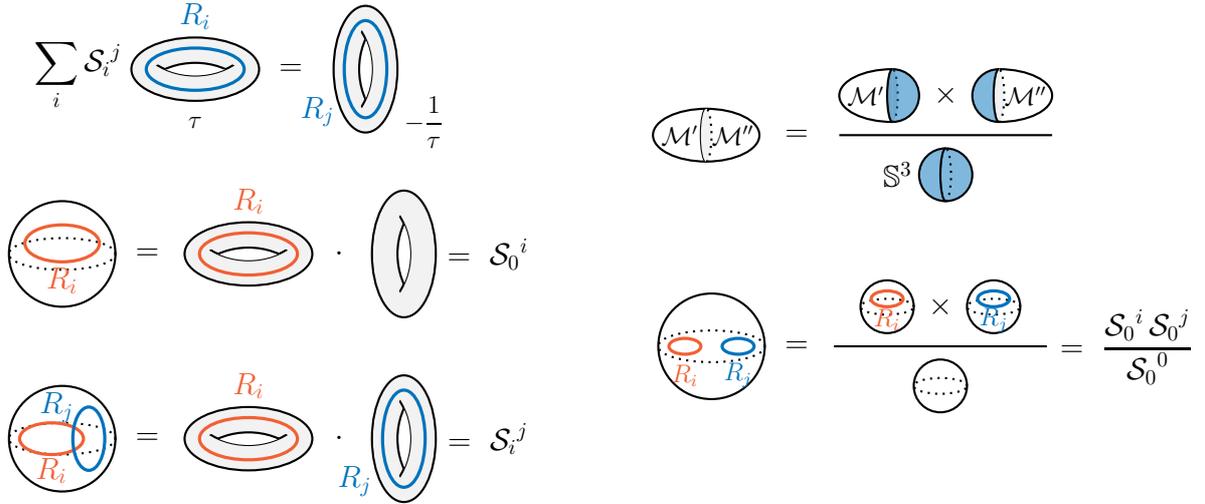

Following the well-known result \cite{Witten:1988hf}, we find that the CFT counterpart of the vacuum partition function of the dual SU$(2)$ Chern-Simons gauge theory on $\BS^3$ is given by 
\begin{align}
    Z_{\text{SU}(2)}\left[\mathbb{S}^3,R_0\right]
    =
    \Smat{0}{0}  \ , \label{S00}
\end{align}
in terms of the $\mathcal{S}$-matrix. We can insert a Wilson loop in the representation $R_j$ on $\mathbb{S}^3$, and the Chern-Simons partition function is computed as
\begin{align}
    Z_{\text{SU}(2)}\left[\mathbb{S}^3,R_j\right]
        =
        \Smat{0}{j}   \, .
\end{align}
Refer to figure \ref{wilsa} for a brief sketch of the derivation via surgery method. 
We can consider a more complicated setup with two Wilson loops in $R_j$ and $R_l$ representations.
When the two Wilson loops are linked, 
the partition function is given by
\begin{align}
Z_{\text{SU}(2)}\left[\mathbb{S}^3,L(R_j,R_l)\right]=\Smat{l}{j}  \, .
\end{align}
When two Wilson lines are not linked (see figure \ref{wilsa} again), we have 
\ba
Z_{\text{SU}(2)}\left[\mathbb{S}^3,R_j,R_l\right]=\frac{\Smat{0}{j}\, \Smat{0}{l}}{\Smat{0}{0}} \, .
\ea

Next, we take the $k\to -2$ limit 
\begin{align}\label{cricialN2}
k = - 2 +  \i\,\frac{ 6  }{c^{(g)}} + \mathcal{O} \left(\frac{1}{(c^{(g)})^{2}}\right) \ ,
\end{align}
with large $c^{(g)}(\in \mathbb{R})$ in order to compare with classical gravity such that the central charge is related to the radius $L$ of $\mathbb{S}^3$ via \cite{Banados:1998tb}
\ba
c^{(g)}=\frac{3L}{2G_N} \, .  \label{BHen}
\ea
This indeed corresponds to the large central charge of the $\SU(2)$ WZW model as in (\ref{centtwo}).
Moreover, the (chiral) conformal weight of a primary state in the representation $R_j$ becomes
\begin{align} \label{hjsu2}
h_j = \frac{j(j+1)}{k+2}\simeq - \i\, \frac{ c^{(g)} j (j+1)}{6} \equiv \i\, h^{(g)}_j \ ,
\end{align}
with $h^{(g)}_j \in \mathbb{R}$ at the limit.  We also call the total conformal weight as $\Delta_j=2h_j=\i \, \Delta^{(g)}$. As usual in the AdS/CFT correspondence, the energy $E_j$ of the object dual to a primary state (or Wilson loop) in the representation $R_j$ is related to conformal dimension $\Delta_j^{(g)}$ as 
\begin{align}
L\, E_j=\Delta^{(g)}_j = 2h^{(g)}_j \, . \label{enegyrel}
\end{align}
It is useful to note the relation
\begin{align}
(2j+1)^2\simeq 1-8\,G_N E_j=1-\frac{24\,h^{(g)}_j}{c^{(g)}} \, . \label{E2j}
\end{align}

Now let us work out the behavior of partition functions in our limit  $k\to -2$. The $\SU(2)$ WZW (non-chiral) CFT can be regarded as a double copy of the $\SU(2)$ Chern-Simons gauge theories as in 
(\ref{CS_WZW}) and (\ref{CS_WZWA}).
Thus, in our limit  $k\to -2$, the vacuum partition function is evaluated as follows:
\begin{align} \label{vpartition}
Z_\text{CFT}[R_0]=\left|Z_{\text{SU}(2)}[\mathbb{S}^3 , R_0]\right|^2  = \frac{c^{(g)}}{3}\sinh^2 \left( \frac{\pi c^{(g)}}{6} \right) \simeq \frac{c^{(g)}}{12}\, \exp\left[\frac{\pi c^{(g)}}{3}\right]\, . 
\end{align}
With the insertion of a Wilson loop in the representation $R_j$, the partition function becomes
\begin{align} \label{singleg}
Z_\text{CFT}[ R_j] = \left|Z_{\text{SU}(2)}[\mathbb{S}^3 , R_j]\right|^2 
 &\simeq \frac{c^{(g)}}{12}\,\exp\left[\frac{\pi c^{(g)}}{3}\s{1-8 G_N E_j}\right] \, , 
\end{align}
where we have used the relation (\ref{E2j}).
Similarly, the 
partition function on $\mathbb{S}^3$ corresponding to the one with two linked Wilson loops can be evaluated as follows:
\begin{align}\label{twowa}
    \begin{aligned}
     Z_\text{CFT}\left[L(R_j,R_l)\right] 
        &= 
        \left|Z_{\text{SU}(2)}\left[\mathbb{S}^3 , L (R_j,R_l)\right]\right|^2 \\
        &\simeq 
        \frac{c^{(g)}}{12}\,\exp\left[\frac{\pi c^{(g)}}{3}\s{1-8G_N E_j}\s{1-8 G_N E_l}\right] \, .
    \end{aligned}
\end{align}
When the two Wilson lines are not linked we find
\begin{align}
\begin{aligned}\label{twoab}
 Z_\text{CFT} [R_j,R_l]
&=\frac{c^{(g)}}{3}\,\frac{\sinh^2\left[\frac{\pi c^{(g)}}{6}\,\s{1-8 G_N E_j}
\right]\sinh^2\!\left[\frac{\pi c^{(g)}}{6}\s{1-8G_N E_l}\right]}
{\sinh^2\left[\frac{\pi c^{(g)}}{6}\right]} \\
& \simeq \frac{c^{(g)}}{12}\,\exp\left[\frac{\pi c^{(g)}}{3}\left(\s{1-8G_N E_j}+\s{1-8G_N E_l}-1\right)\right] \, .  
\end{aligned}
\end{align}

\subsection{Gravity calculation}
\label{sec:gravitysu2}

Here we perform gravity calculations and reproduce the CFT results obtained in the previous subsection. We define the gravity partition function by
\begin{align}
Z_\text{G} = e^{- I_\text{G}} \, , \qquad I_\text{G} = - \frac{1}{16 \pi G_N} \int\d^3 x\, \sqrt{g} \left(R - 2 \Lambda\right) \,  ,
\end{align}
where the cosmological constant is related to the $\mathbb{S}^3$ radius $L$ as $\Lambda = L^{-2}$.
On $\mathbb{S}^3$, the Ricci curvature is given by $R = 6 \Lambda$. The solution to the Einstein equation in three dimensions with the positive cosmological constant is locally the same as $\mathbb{S}^3$. Thus, the action can be written as
\begin{align}
 I_\text{G} = - \frac{1}{4 \pi L^2 G_N} \int\d^3 x\, \sqrt{g} \, .
\end{align}

We begin with the partition function on $\mathbb{S}^3$ without any excitation. The geometry can be described by a hypersurface
\begin{align}
X_1^2 + X_2^2 + X_3^2 + X_4^2 = L^2 \ ,
\end{align}
in $\mathbb{R}^4$ with the metric $\d s^2 = \sum_{j=1}^4 \d X_j^2$.
A convenient parametrization is given by
\begin{align}
\begin{aligned}
X_1=L \cos\theta \cos \tau \, ,\qquad
X_2=L \cos\theta \sin \tau \, , \\
X_3=L \sin\theta \cos \phi \, ,\qquad
X_4=L \sin\theta \sin \phi \, ,
\end{aligned}
\end{align}
which leads to the metric
\begin{align}
\d s^2 = L^2 \left[\d \theta^2 + \cos ^2 \theta\, \d \tau ^2 + \sin^2 \theta\, \d\phi^2 \right] \, . \label{S3metric}
\end{align}
Here we take $0 \leq \theta \leq \pi/2$ and identify $\tau \sim \tau + 2\pi$, $\phi \sim \phi + 2 \pi$.
Then, we can evaluate the action as 
\begin{align}
 I_\text{G} = - \frac{\pi L}{2 G_N}  = - \frac{\pi c^{(g)}}{3} \, . \label{vgravity}
\end{align}
Thus, the gravity partition function $Z_\text{G} = \exp (- I_\text{G})$ reproduces \eqref{vpartition} in the large $c^{(g)}$ limit as expected. Note that $-I_\text{G}$ describes the entropy of three-dimensional de Sitter space.

We then move to the case with an extra insertion of Wilson loop in the representation $R_j$.
In the dual gravity side, it corresponds to the geometry with  excitation energy $E_j$ given in \eqref{enegyrel}. 
The geometry is the Euclidean de Sitter black hole solution with the metric
\begin{align}
    \d s^2=L^2\left[(1-8G_N E_j -r^2)\,\d\tau^2+\frac{\d r^2}{1-8G_N E_j -r^2}+r^2 \d \phi^2\right] \, . \label{dSBH}
\end{align}
In higher dimensions, there are two horizons corresponding to the black hole horizon and cosmological horizon. However, in three dimensions, there is only one horizon at $r=\s{1-8G_N E_j}$.
If we assume that the geometry is smooth at the horizon, then the periodicity of Euclidean time direction $\tau$ is fixed as
\begin{align}
\tau\sim \tau+\frac{2\pi}{\s{1-8G_N E_j}} \, .
\end{align}
The angular coordinate $\phi$ satisfies the periodicity condition $\phi\sim \phi+2\pi$ and there is a conical singularity at $r=0$ with the deficit angle $\delta = 2 \pi (1 - \sqrt{1 - 8 G_N E_j})$.
From the area of the horizon, we can read off the entropy as
\begin{align}
S_\text{dS\,BH}=\frac{2\pi \s{1-8G_N E_j}}{4G_N}=\frac{\pi}{3}c^{(g)}\s{1-8G_N E_j} \ .\label{BHent}
\end{align}
We may perform a coordinate transformation as
\begin{align}
r=\s{1-8G_N E_j}\,\sin\theta \, , \qquad \tau ' = \s{1-8G_N E_j}\,\tau
\end{align}
with $0\leq \theta\leq \pi/2$ and the periodicity $\tau ' \sim \tau ' + 2 \pi$.
The metric becomes
\begin{align}
\d s^2 = L^2 \left[ \d\theta^2 + \cos^2\theta \,\d\tau ' {}^2 + (1 - 8G_N E_j) \sin^2\theta\, \d\phi^2\right] \ .
\label{metsph}
\end{align}
Then we can evaluate the action for this solution and obtain 
\begin{align}
I_\text{G} =-\frac{\pi}{3}c^{(g)}\s{1-8G_N E_j} \, .
\end{align}
The gravity partition function $Z_\text{G}=\exp (-I_\text{G})$ perfectly agrees with the previous result \eqref{singleg}.

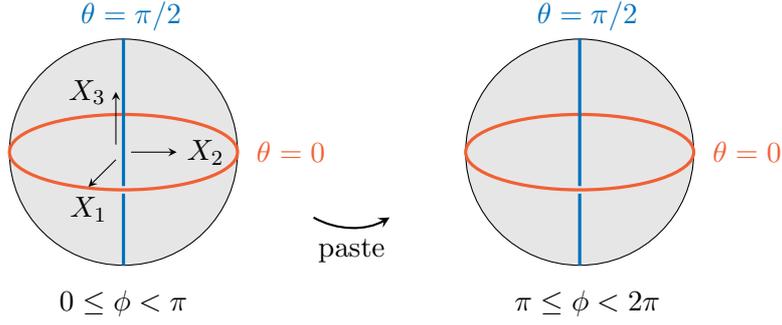
\begin{figure}
  \centering
    \begin{tikzpicture}[>=stealth]
        \draw[fill=gray!20] (-2,0) circle (1.5cm);
        \draw[RedOrange, very thick] (-2, 0) ellipse (1.5cm and 0.5cm);
        \node[RedOrange] at (0.2, 0) {$\theta=0$};
        \draw[-, RoyalBlue, very thick] (-2, 1.5) -- (-2, -0.45);
        \draw[-, RoyalBlue, very thick] (-2, -0.55) -- (-2, -1.5);
        \node[RoyalBlue] at (-1.9, 1.8) {$\theta=\pi/2$};
        \draw[->] (-2.1, -0.1) --+ (-0.36, -0.36) node[below] {$X_1$};
        \draw[->] (-1.9, 0) --+ (0.6, 0) node[right] {$X_2$};
        \draw[->] (-2.1, 0.1) --+ (0, 0.7) node[left] {$X_3$};
        \node at (-2, -2) {$0 \leq \phi < \pi$};
        
        \begin{scope}[xshift=2cm]
            \draw[fill=gray!20] (2,0) circle (1.5cm);
            \draw[RedOrange, very thick] (2, 0) ellipse (1.5cm and 0.5cm);
            \node[RedOrange] at (4.2, 0) {$\theta=0$};
            \draw[-, RoyalBlue, very thick] (2, 1.5) -- (2, -0.45);
            \draw[-, RoyalBlue, very thick] (2, -0.55) -- (2, -1.5);
            \node[RoyalBlue] at (2.1, 1.8) {$\theta=\pi/2$};
            \node at (2.1, -2) {$\pi \leq \phi < 2\pi$};
        \end{scope}
        
        \begin{scope}[xshift=1cm]
            \draw[->, thick] (240:1) arc (240:300:1) node[midway, below] {paste};
        \end{scope}
    \end{tikzpicture}
\caption{The North (left) and South (right) hemisphere with two linked Wilson lines (orange and blue).}
\label{wilsc}
\end{figure}

We then move to a more complicated example of geometry, which corresponds to the insertion of two Wilson loops in the representations $R_j$ and $R_l$.
Recall that the geometry with the metric  \eqref{metsph} corresponds to the one with the conical defect at $\theta =0$ with deficit angle $\delta_j = 2 \pi (1 - \sqrt{1 - 8 G_N E_j})$. We can also include another conical defect at $\theta = \pi/2$ by changing the periodicity as $\tau '' \sim \tau''  + 2 \pi \sqrt{1 - 8 G_N E_l}$ or equivalently  performing a coordinate transformation
\begin{align}
\tau '' = \sqrt{1 - 8 G_N E_l }\ \tau '  \, , \qquad  \tau ' \sim \tau ' + 2\pi \, .
\end{align}
The deficit angle at $\theta = \pi/2$ is  $\delta_l=2\pi \left( 1 -\s{1-8 G_N E_l}\right)$ and 
the metric now reads 
\begin{align}
\d s^2 = L^2 \left[ \d\theta^2 + (1 - 8G_N E_l) \cos^2\theta\, \d\tau ' {}^2 + (1 - 8G_N E_j) \sin^2\theta\, \d\phi^2\right] \, .
\label{N2conical}
\end{align}
The second conical singularity corresponds to the second Wilson loop at $\theta=\pi /2$, which is linked to the first one, see figure \ref{wilsc}.
The gravity action for this geometry can be found as
\begin{align}
I_\text{G}=-\frac{\pi}{3}c^{(g)}\s{1-8G_N E_j}\,\s{1-8G_N E_l} \, .  \label{doubleW}
\end{align}
This again agrees with the previous result in \eqref{twowa}.

Finally, we examine the geometry corresponding to the insertion of two unlinked Wilson lines in the representations $R_j$, $R_l$. Let us recall that the energy and the label of representation are related as 
\begin{align}
\sqrt{1 - 8 G_N E_i} = 2i + 1 \label{E2r}
\end{align}
as in \eqref{E2j}. This relation implies that
\begin{align}
 \sqrt{1 - 8 G_N E_{j + l}}  =  2(j + l) + 1
 = (2j +1) + (2l+1) - 1 = \sqrt{1 - 8 G_N E_j}  +  \sqrt{1 - 8 G_N E_l} - 1 \, .
\end{align}
Therefore, the corresponding geometry is the Euclidean de Sitter black hole with the metric \eqref{dSBH} but the energy is now $E_{j+l}$ defined by \eqref{E2r} with $i = j +l$. The gravity action is 
\begin{align}
I_\text{G}=-\frac{\pi}{3}c^{(g)}\s{1-8G_N E_{j+l}} =-\frac{\pi}{3}c^{(g)} \left( \s{1-8G_N E_{j}}  +  \s{1-8G_N E_{l}}  -1 \right) \, , 
\end{align}
which reproduces the previous result in \eqref{twoab}.

\subsection{De Sitter entropy as topological entanglement entropy}
It is intriguing to pose here to study entanglement entropy in the gravity theory on $\mathbb{S}^3$ and its CFT dual.
We choose a subsystem disk $A$ on the boundary $\mathbb{S}^2$ of a three-dimensional ball $\BB^3$, on which the wave functional of 3d gravity is defined. We write the boundary of $A$ as $\Gamma_A$, which is a circle. In the replica calculation of entanglement entropy, we introduce a  cut along $\Gamma_A$ on $\mathbb{S}^3$ and take its $n$-folds to obtain $\mbox{Tr}[(\rho_A)^n]$, where $\rho_A$ is a reduced density matrix
obtained by tracing out the complement of $A$.
The topological entanglement entropy \cite{Kitaev:2005dm,Levin:2006zz} takes the following form  in the presence of an Wilson line dual to a massive excitation with energy $E$
\ba
S_A=\log |\Smat{0}{j}|^2=\frac{\pi}{3}c^{(g)}\s{1-8G_NE} \, ,
\label{topEE}
\ea
which coincides with the BH entropy (\ref{BHent}). Note that in the replica partition function does not depend on the parameter $n$ as the $n$-fold cover of $\BS^3$ is topologically $\BS^3$.
Refer to \cite{McGough:2013gka}
for an analogous relation in black holes in AdS$_3/$CFT$_2$. 

Alternatively, we may regard the conical deficit along $\Gamma_A$ as a Wilson loop $W(\Gamma_A)$.
Given the Chern-Simons/WZW correspondence, this loop is seen as the worldline of the twist operator in the dual CFT of the energy
\ba
E_n\, L=\frac{c}{12}\left( n -\frac{1}{n}\right) \, .
\ea
If there is a Wilson line excitation of energy $E$ which links $\Gamma_A$, 
the entanglement entropy can also be calculated by employing the formula (\ref{doubleW}) for the two linked Wilson loops:
\ba
S_A=-\frac{\de}{\de n}\log \,\la\, W(\Gamma_A)\,\lb\bigg|_{n=1} =\frac{\pi}{3}c^{(g)}\,\s{1-8G_NE} \, ,
\ea
reproducing the same value as \eqref{topEE}.

We expect that we can extend the above relation to topological pseudo entropy \cite{Nakata:2020luh,Nishioka:2021cxe},  which generalizes the entanglement entropy such that the entropy depends on both the initial and final states.

\section{Higher-spin generalization}
\label{sec:hs}

In this section, we extend the analysis in the previous section by considering the duality for generic $N$, summarized as (\ref{dualityN}).
There is not much difference in the CFT computation except that expressions become more group theoretical.
In the gravity side, we use the Chern-Simons formulation of higher-spin gravity, which gives alternative viewpoints on the pure gravity case with $N=2$.

\subsection{CFT description}

In this subsection, we obtain the partition functions on $\mathbb{S}^3$ in the presence of Wilson loops from the $\mathcal{S}$-matrix for the modular transformation of characters of SU$(N)$ current algebra.
We consider the SU$(N)$ WZW model with the central charge given in \eqref{centNa}, and we are interested in the $k\to -N$ limit: 
\ba
    k = -N +\i\,\frac{N(N^2-1)}{c^{(g)}}+O\left(\frac{1}{(c^{(g)})^{2}}\right) \, ,   \label{limit0}
\ea
where $c^{(g)}$ is related to the radius of $\mathbb{S}^3$ as in \eqref{centNa} 
and thus the classical higher-spin limit is $c^{(g)}\to \infty$.

It will be convenient to introduce a Lie-algebraic notation for the comparison with computations in higher-spin gravity.
We introduce $N$-dimensional basis $e_i$ $(i=1,2,\ldots,N)$ associated with the inner product $(e_i , e_j) = \delta_{i,j}$. The simple roots of SU$(N)$ are given by $\alpha_i = e_i - e_{i+1}$ with $i=1,2,\ldots , N-1$ and the fundamental weights are
\begin{align}
    \omega_j = \sum_{l =1}^j e_l - \frac{j}{N} \sum_{l=1}^N e_l \ ,
\end{align}
with $j=1,2,\ldots , N-1$. The Weyl vector is then defined as
\begin{align}
    \rho = \sum_{j=1}^{N-1} \omega_j = \sum_{j=1}^N \left( \frac{N+1}{2} - j \right) e_j \, . \label{rho}
\end{align}
The highest weight state can be labeled by a Young diagram $\mu$, which may be expressed as
\begin{align}
    \mu = \sum_{l=1}^{N-1} \lambda_l\, \omega_l =  \sum_{j=1}^{N} \mu_j\, e_j  \, .
\end{align}
Here $\lambda_l$ corresponds to the Dynkin label.%
\footnote{Here we assumed that the labels $\lambda_l$ $(l=1,2,\ldots,N-1)$ are non-negative integer. We will see below that the states are dual to conical defects satisfying the trivial holonomy conditions \eqref{trivilahol}. In this paper, we discuss only the leading order in $1/c^{(g)}$, and at least in this case more generic states with non-integer $\lambda_l$ (or conical defects not satisfying the trivial holonomy conditions) are allowed. We shall discuss this issue in more details in section \ref{sec:spectrum}.}
When the Young diagram $\mu$ has $r_j$ box in the $j$-th row,
then $\lambda_l = r_l - r_{l-1}$ and $\mu_j = r_j - |\mu|/N$. Here $|\mu|$ represents the number of total box in the Young diagram $\mu$. With the terminology, the central charge and the conformal weight of primary operator labeled by $\mu$ are
\begin{align}
    c = \frac{12 k\, (\rho ,\rho )}{N (N + k)} \, , \qquad h_\mu = \frac{C_2(\mu)}{N +k} \, , \qquad C_2 (\mu) = \frac12\, (\mu , \mu+ 2 \rho) \, . \label{hsh}
\end{align}
Here we use
\begin{align}
    (\rho,\rho)= \frac{1}{12} N (N^2 -1)\ ,
\end{align}
and denote the eigenvalue of the quadratic Casimir operator by 
$C_2 (\mu)$.

The character and modular $\mathcal{S}$-matrix can be defined similarly to the SU$(2)$ case as
\begin{align}
    \chi_\mu = \text{Tr}_{R_\mu} \left[e^{2 \pi\, \i\, \tau \left(L_0-\frac{c}{24}\right)}\right] \, , \qquad \chi_\mu (-1/\tau) = \sum_{\nu } \Smat{\mu}{\nu}\, \chi_\nu (\tau) \, .
\end{align}
Here $R_\mu$ is the representation with the highest weight labeled by the Young diagram $\mu$.
The expression of the $\mathcal{S}$-matrix may be found, e.g., in  \cite{DiFrancesco:1997nk} as
\begin{align}
\Smat{\mu}{\nu}=K\sum_{w\in W}\ep(w)\,\exp\left[-\frac{2\pi\, \i}{k+N}(w(\mu+\rho),\,\nu+\rho)\right] \, , \label{SN}
\end{align}
where $W$ is the Weyl group of SU$(N)$ and $\epsilon (w) = \pm 1$ is associated with the group element. Moreover, $K$ is a constant fixed by the unitary constraint $\mathcal{S}\, \mathcal{S}^\dagger=1$.
In the $k\to -N$ limit of \eqref{limit0}, only a term dominates in the sum over the elements of Weyl group.%
\footnote{An element of Weyl group for SU$(N)$ permutes the $N$-dimensional basis $e_i$. It is convenient to introduce the charge conjugation $\mu^* \equiv - w_0 \mu$, where $w_0$ exchanges $e_i$ by $e_{N-i}$ for all $i \leq \lfloor N/2 \rfloor $. Note that the Weyl vector is self-conjugate as $\rho = - w_0 \rho$. We can see that the dominant contribution at the critical level comes from the term with $w = w_0$. For instance, $\mu^*_i + \rho_i > \mu^*_j + \rho_j$ and $\nu_i + \rho_i > \nu_j + \rho_j$ for $i > j$, and hence the exchange of $\mu^*_i + \rho_i$ and $\mu^*_j + \rho_j$ makes $(\mu^* + \rho , \nu + \rho)$ smaller.}
Let us denote the central charge $c=\i \, c^{(g)}$ as in \eqref{centNa}.
For the case with two linked Wilson loops in the representations $R_{\mu^*}, R_\nu$, we have
\begin{align}
Z_\text{CFT} \left[L (R_{\mu^*} , R_\nu ) \right] = \left|\Smat{\mu^*}{\nu} \right|^2 \simeq \exp\left[\frac{\pi}{3} c^{(g)} \frac{(\mu + \rho , \nu + \rho )}{(\rho , \rho)}\right] \, . 
\label{cft}
\end{align}
Here we ignored the coefficient in front of the exponential as we are only interested in the large central charge limit. 
For example, the vacuum partition function is
\begin{align}
Z_\text{CFT} [R_0] =  \left|\Smat{0}{0} \right|^2 \simeq \exp\left[\frac{\pi}{3} c^{(g)}\right] \, , \label{vpartitionN}
\end{align}
which is the same as \eqref{vpartition} for the $N=2$ case. The partition function with a Wilson loop in the representation $R_\mu$ is given by
\begin{align}
 Z_\text{CFT}[ R_\mu ]  = \left|\Smat{0}{\mu} \right|^2 \simeq \exp\left[\frac{\pi}{3} c^{(g)} \frac{( \rho , \mu + \rho )}{(\rho , \rho)}\right] \, . \label{WilsonN}
\end{align}
Moreover, the unlinked case reads
\begin{align}
    \begin{aligned}
     Z_\text{CFT}\left[R_\mu, R_\nu\right] &\simeq \left| \frac{\Smat{0}{\mu} \Smat{0}{\nu}}{\Smat{0}{0}}\right|^2\\
     &\simeq \exp\left[\frac{\pi}{3} c^{(g)} \left( \frac{( \rho , \mu  + \rho )}{(\rho , \rho)} +  \frac{( \rho , \nu  + \rho )}{(\rho , \rho)}  - 1  \right) \right]\\
     &=
    \exp\left[\frac{\pi}{3} c^{(g)} \frac{( \rho , \mu + \nu + \rho )}{(\rho , \rho)}\right]
\,  ,
    \end{aligned}
\end{align}
which is the same as the case with one Wilson line in the representation $R_{\mu + \nu}$.

\subsection{Conical defect geometry}
Now we move on to the higher-spin gravity calculation.
Here we consider higher-spin gravity on $\mathbb{S}^3$ described by $\text{SU}(N) \times \text{SU}(N)$
Chern-Simons theory and construct a conical defect geometry.
The action of the Chern-Simons theory is 
\begin{align}
&	I_\text{CSG} = I_\text{CS} [A] - I_\text{CS} \left[\bar A\right] + \frac{\kappa}{4 \pi} \int_{\partial \mathcal{M}} \text{Tr} \left(A \wedge \bar A\right) \, , \label{CSaction0}
 \\
&	I_\text{CS}[A]  =  - \frac{\kappa}{4 \pi} \int_{\mathcal{M}} \text{Tr} \left( A \wedge \d A + \frac{2}{3}\, A \wedge A \wedge A \right) \, . \label{CSaction}
\end{align}
The boundary term is included since we will use a slightly singular coordinate system. In the next subsection, we will see that there is a gauge choice such that the boundary term does not contribute to the on-shell action.
The gauge fields are defined as
\begin{align}
	A = A^a_\mu\, L_a\, \d x^\mu  \, , \qquad \bar A = \bar A^a_\mu\, \bar L_a\, \d x^\mu \, ,
\end{align}
where we use the convention of \cite{Castro:2020smu} for $\bar L_a$.
For the definition of higher-spin gravity, it would be important to identify the gravitational sector SU$(2)$ inside SU$(N)$. We use so-called the principal embedding of SU$(2)$,%
\footnote{In this paper, we only consider the principal embedding of SU$(2)$, see, e.g., \cite{Anninos:2020hfj} for other embeddings. Conical defects for higher-spin AdS$_3$ gravity with other embeddings have been analyzed in \cite{Castro:2011iw}. Moreover, conical defects are identified with states in dual CFTs for other higher-spin AdS/CFT correspondences, e.g., in \cite{Hikida:2012eu} for $\mathcal{N}=2$ higher-spin holography of \cite{Creutzig:2011fe} and in \cite{Creutzig:2019qos} for matrix extended higher-spin holography of \cite{Creutzig:2013tja,Creutzig:2018pts}.}
which is generated by 
\begin{align}
  L_3 = \sum_{i=1}^N \rho_i\, e_{i,i} \ ,
\end{align}
and $L_1,L_2$ satisfying $[L_i , L_j] = \frac{\i}{2} \epsilon_{ijk}\, L_k$. 
Here $e_{i,j}$ are $N \times N$ matrices with elements:
\begin{align}
    (e_{i,j})_k^{~l} = \delta_{i,k}\, \delta_j^{~l} \ .
\end{align}
The normalization of generators is given by
\begin{align}
 \text{Tr} (L_i\, L_j) = \epsilon_N\, \delta_{i,j} \,  , \qquad \epsilon_N = (\rho , \rho) = \frac{1}{12}N (N^2-1) \, , 
\end{align}
see \eqref{epsilonN}.
The Chern-Simons level is related to the Newton constant as
\begin{align} \label{hsk2}
	\kappa = \frac{L}{8 G_N \epsilon_N}
\end{align}
in this convention.

The gauge fields are linear combinations of higher-spin generalizations of dreibein and spin connections.
In particular, the metric can be read off as 
\begin{align}\label{gauge_metric}
	g_{\mu\nu} = - \frac{L^2}{4 \epsilon_N}\, \text{Tr} \left[ (A_\mu - \bar A_\mu) (A_\nu - \bar A_\nu)\right] \, .
\end{align}
The gauge configuration corresponding to $\mathbb{S}^3$ may be given by
\begin{align}
	A = b^{-1}\, a\, b + b^{-1}\, \d b  \ , \qquad \bar A = b\, \bar a\, b^{-1} + b\, \d b^{-1} \ ,  \label{flat}
\end{align}
where
\begin{align}
	a = \i \, L_1 (d \phi + d \tau) \,  , \qquad \bar a = \i \, L_{1} (d \phi - d \tau) \,  , \qquad b = e^{\i\, \theta L_3 } \,   .
\end{align}
We can obtain the metric in \eqref{S3metric} from the formula \eqref{gauge_metric}.

A solution to the equations of motion obtained from the Chern-Simons action is given by a flat connection, which can be put into the form \eqref{flat} by a gauge fixing.  Here we focus on the solutions with constant $a ,\bar a$. The metric is actually transformed under gauge transformation, therefore it has not so significant meaning.
Here we construct ``conical defect geometry'' in the sense of  \cite{Castro:2011iw}, i.e.,  a geometry described by a gauge configuration with a trivial holonomy.
We require the triviality condition for the holonomies along $\phi$-cycle around $\theta =0$ and along $\tau$-cycle around $\theta =\pi/2$. These holonomies are defined by
\begin{align}
	\text{Hol}_\phi (A) = \mathcal{P} \exp \left( \oint_{\theta=0} A \right) \,  ,  \qquad
	\text{Hol}_\tau (A) = \mathcal{P} \exp \left( \oint_{\theta=\pi/2} A \right)  \, , \label{trivilahol}
\end{align}
where $\mathcal{P}$ represents the path ordering.
We mean by a trivial holonomy here that it is proportional to $\pm \mathbbm{1}_{N \times N}$ for $N$ even and  $\mathbbm{1}_{N \times N}$ for $N$ odd.%
\footnote{Precisely speaking, we consider the gauge group (SU$(N) \times$SU$(N)$)$/\mathbb{Z}_2$ for even $N$, and the minus sign comes from the $\mathbb{Z}_2$, see \cite{Castro:2011iw} as well.}
Such geometry may be realized by gauge configuration with
\begin{align}\label{gauge_conical}
	a = - \sum_{i=1}^{\lfloor\frac{N}{2}\rfloor} B_{2i-1}^{(1)}(1, 1)(n_i\, \d \phi + \tilde n_i\, \d \tau ) \,  ,  \qquad
\bar a = - \sum_{i=1}^{\lfloor\frac{N}{2}\rfloor} B_{2i-1}^{(1)}(1, 1)(n_i\, \d \phi - \tilde n_i\, \d \tau )  \, , 
\end{align}
where $n_i , \tilde n_i \in \mathbbm{Z}$ for odd $N$ and $n_i  , \tilde n_i \in \mathbbm{Z}$ or  $n_i , \tilde n_i \in \mathbbm{Z} + 1/2$ for even $N$. 
Here we define a matrix $B_k^{(l)}$ by
\begin{align}
    \left[ B_k^{(l)}(x,y)\right]_{ij} \equiv x\,\delta_{i,k}\,\delta_{j,k+l} - y\,\delta_{i,k+l}\,\delta_{j,k} \, ,
\end{align}
which satisfies the relation:
\begin{align}
    e^{\rho L_3}\,B_k^{(l)}(x,y)\,e^{-\rho L_3} = B_k^{(l)}\left(e^{l\rho} x,e^{-l\rho} y\right) \, .
\end{align}
The metric corresponding to the gauge configuration \eqref{gauge_conical} is
\begin{align}\label{conical_metric}
	\d s^2 = \d\theta^2 +  R_N^2  \cos ^2 \theta\, \d \tau^2 + \tilde R_N^2  \sin ^2 \theta\, \d \phi^2 \ ,
\end{align}
with parameters
\begin{align}
	R_N ^2 =  \sum_{i=1}^{\lfloor\frac{N}{2}\rfloor}   \frac{2 n_i^2 }{\epsilon_N} \, , \qquad	\tilde R_N ^2 =  \sum_{i=1}^{\lfloor\frac{N}{2}\rfloor}   \frac{2 \tilde n_i^2 }{\epsilon_N} \, .
\end{align}
This metric is for a conical geometry with conical deficit angles $2 \pi (1 - R_N)$ around $\tau$-cycle at $\theta = \pi$ and $2 \pi (1 -\tilde R_N)$ around $\phi$-cycle at $\theta = 0$. 
The conical defect geometry reproduces \eqref{N2conical} for $N=2$.
As in the case of Euclidean AdS$_3$ analyzed in \cite{Castro:2011iw}, we consider the solutions with negative deficit angles, i.e., conical surplus geometry.
This means that we set $R_N \geq 1 $ and $\tilde R_N \geq 1 $.
The equalities $R_N = \tilde R_N = 1$ can be realized by choosing $n_i = \tilde n_i = \rho_i$ (see \eqref{rho}), which correspond to $\mathbb{S}^3$-background. 

We proceed to evaluate the action for the gauge configuration. The bulk Chern-Simons action in \eqref{CSaction} vanishes for the ansatz, and the whole contribution comes from the boundary term in \eqref{CSaction0} at $\theta = 0 , \pi/2$:
\begin{align}
	I_\text{CSG} = - \frac{\pi L \sum_{i=1}^{\lfloor\frac{N}{2}\rfloor } n_i\, \tilde n_i}{G_N \epsilon_N } \,  . \label{conicalaction}
\end{align}
Let us change the order of parameters as $n_1 \geq n_2 \ldots $, $\tilde n_1 \geq \tilde n_2 \ldots $  and set
\begin{align}
	n_i = - n_{N+1 - i} \, , \qquad 	\tilde n_i = - \tilde n_{N+1 - i}  \label{condYoung}
\end{align}
for $i > \lfloor\frac{N}{2}\rfloor $. 
From now on, we require $n_i \neq n_j$  and $\tilde n_i \neq \tilde n_j$  for $i \neq j$, which generically corresponds to the diagonalizability of matrix. Then, we could set 
\begin{align}
	n_i =  \mu_i  + \rho_i \ , \qquad \tilde 	n_i = \nu_i  +  \rho_i\ .
\end{align}
Expressing 
\begin{align}
2 \sum_{i=1}^{\lfloor\frac{N}{2}\rfloor} n_i\, \tilde n_i =  \sum_{i=1}^N n_i\, \tilde n_i = (\mu + \rho , \nu + \rho) \ ,
\end{align}
and using $c^{(g)} = 3 L / (4 \epsilon_N G_N) $ as in \eqref{BHen},
we can rewrite \eqref{conicalaction} as
\begin{align}
	I_\text{CSG} = - \frac{\pi c^{(g)} }{3} \frac{( \mu + \rho ,  \nu + \rho)}{(\rho,\rho)}\, . \label{bulk}
\end{align}
This reproduces the partition function on $\mathbb{S}^3$ obtained from the modular $\mathcal{S}$-matrix as in \eqref{cft}. However, note that Young diagrams obtained in this way are not generic ones due to the extra condition \eqref{condYoung}. The same condition actually arises in the case of Lorentzian AdS$_3$ and the condition can be relaxed by moving to Euclidean AdS$_3$ \cite{Castro:2011iw}. We expect that the condition \eqref{condYoung} can be removed by working on Lorentzian dS$_3$.

Here we remark that, in the case where $\mu$ is trivial, $-I_\text{CSG}$ is identical to the de Sitter black hole entropy in the higher-spin gravity and also equals to the topological entanglement entropy as in the case of Einstein gravity (\ref{topEE}):
\ba
S_\text{dS\,BH}=S_A=\log \left|\Smat{0}{\nu}\right|^2=\frac{\pi c^{(g)} }{3} \frac{( \rho , \nu + \rho)}{(\rho,\rho)} \, .
\ea

\subsection{Conical geometry in another gauge}
The gauge configuration \eqref{gauge_conical} for the conical geometry constructed in the previous subsection is singular at $\theta = 0$ and $\theta = \pi/2$ and the on-shell action has a contribution purely from the boundary term.
This is conceptually confusing since there are no boundaries at $\theta = 0, \pi/2$ on $\BS^3$ but the boundary term plays a physically important role.
We will show that this issue can be reconciled by moving to another gauge configuration that is gauge-equivalent to the original one \eqref{gauge_conical}.

To this end, we note that the gauge configuration \eqref{gauge_conical} describing the conical geometry can be written as
\begin{align}
    a = h^{-1}\,\d h \ , \qquad \bar a = \bar h \, \d \bar h^{-1} \ ,
\end{align}
where
\begin{align}
        h = \prod_{i=1}^{\lfloor\frac{N}{2}\rfloor}\exp\left[ - B_{2i-1}^{(1)}(1,1)\,(n_i\,\phi + \tilde n_i\,\tau)\right] \, , \qquad
        \bar h = \prod_{i=1}^{\lfloor\frac{N}{2}\rfloor}\exp\left[ B_{2i-1}^{(1)}(1,1)\,(n_i\,\phi - \tilde n_i\,\tau)\right] \, .
\end{align}
It follows from \eqref{flat} that the gauge configuration is written as
\begin{align}
        A = (h\,b)^{-1} \d (h\,b) \, ,\qquad
        \bar A = (b\,\bar h)\,\d (b\,\bar h)^{-1} \, ,
\end{align}
or more explicitly
\begin{align}
    \begin{aligned}
        A 
            &=
            \i\,L_3\,\d r - \sum_{i=1}^{\lfloor \frac{N}{2}\rfloor}\,B_{2i-1}^{(1)}\left( e^{-\i r},e^{\i r}\right)\,(n_i\,\d\phi + \tilde n_i\,\d\tau) \, , \\
        \bar A 
            &=
            -\i\,L_3\,\d r - \sum_{i=1}^{\lfloor \frac{N}{2}\rfloor}\,B_{2i-1}^{(1)}\left( e^{\i r},e^{-\i r}\right)\,(n_i\,\d\phi - \tilde n_i\,\d\tau) \, .
    \end{aligned}
\end{align}

The configuration \eqref{flat} still has a residual gauge symmetry that fixes the metric \eqref{gauge_metric}:
\begin{align}
        A \to g^{-1} A\, g + g^{-1}\d g \, , \qquad
        \bar A \to g^{-1} \bar A\, g + g^{-1}\d g \, .
\end{align}
Performing the residual gauge transformation with $g = b\,\bar h$ results in the following gauge configuration:
\begin{align}
    A = g_\text{con}^{-1}\, \d g_\text{con} \, , \qquad \bar A = 0 \, ,
\end{align}
where $g_\text{con} \equiv h\,b^2\, \bar h$.
In this gauge, the chiral part of the gauge field becomes
\begin{align}
    \begin{aligned}
    g\,A\,g^{-1}
        =
        2\,\i\,L_3\,\d r 
        & + 
        2\,\i \sum_{i=1}^{\lfloor \frac{N}{2}\rfloor}\,B_{2i-1}^{(1)}\left( 1, -1\right)\, n_i\,\sin r\,\d\phi 
        - 2 \sum_{i=1}^{\lfloor \frac{N}{2}\rfloor}\,B_{2i-1}^{(1)}\left( 1, 1\right)\, \tilde n_i\,\cos r\,\d\tau \, .
    \end{aligned}
\end{align}
The conical metric \eqref{conical_metric} can be reproduced by substituting this gauge configuration to \eqref{gauge_metric}.

With this gauge choice, the on-shell action \eqref{CSaction0} has a contribution from purely the bulk part:
\begin{align}
    \begin{aligned}
        I_\text{CSG}
            = 
            - \frac{\kappa}{12\pi}\,\int_{\mathbb{S}^3} \text{Tr}\, (A^3) 
            =
            - 8\pi \kappa \sum_{i=1}^{\lfloor \frac{N}{2}\rfloor}n_i\, \tilde n_i \, ,
    \end{aligned}
\end{align}
which agrees with \eqref{conicalaction}. This provides an alternative derivation which does not need the boundary term contribution.

\section{Towards extension to Lorentzian \texorpdfstring{dS$_3$}{dS3}}
\label{sec:2pt}

In the previous sections, we have considered the Euclidean version of the duality, which is interpreted as the relation between the squared norm of the Hartle-Hawking wave functional \eqref{square} and the modular $\mathcal{S}$-matrix of the WZW model. In this section, we consider the original relation \eqref{dsMaldacena} between the Hartle-Hawking wave functional itself and the CFT on the future boundary of the Lorentzian dS$_3$. Since the geometry of Lorentzian dS$_3$ evolves from the initial state constructed by the Euclidean path integral as figure \ref{dSCFTfig}, quantities computed in the boundary CFT are expected to include contributions from the Euclidean part. We check this expectation by evaluating various quantities such as partition function, two-point function, and entanglement entropy.

Firstly, we evaluate the classical limit of the partition function in the Liouville/Toda CFT and show that the result includes a contribution from the Euclidean part of geometry, which is expected to be a modular $\mathcal{S}$-matrix element $\left|\Smat{0}{0}\right|$ from the discussions on the Euclidean version of the duality. 
Secondly, we evaluate the two-point functions by calculating the geodesic length connecting a pair of points of operators. This is mathematically equivalent to computing a Wilson line in the Chern-Simons description of three-dimensional gravity as we show in the appendix \ref{app:Wilsonline}, which is a 
de Sitter analogue of the argument for the AdS$_3/$CFT$_2$ case presented in \cite{Ammon:2013hba}. After that, we show that this gravitational computation can be reproduced from the direct CFT calculation of the two-point functions under a special treatment of the UV cutoff, which is supported from the CFT dual of the hemisphere.
Furthermore, we calculate the holographic entanglement entropy in dS$_3/$CFT$_2$, which is proportional to the geodesic length in our setup, and confirm that the same result can be obtained from the codimension-two holography \cite{Akal:2020wfl}, combined with the brane-world model \cite{Karch:2000ct}.

\subsection{Partition function in Liouville/Toda CFT}

In this subsection, we would like to evaluate the partition function of the boundary CFT on the future boundary $\BS^2$ of Lorentzian dS$_3$. 
We employ the description by Liouville/Toda CFT with $b \to 0$, which should lead to the same results from the description by WZW model with $k \to -N$ as mentioned in the introduction. 
Notations and properties of the Liouville/Toda CFT are summarized in appendix \ref{app:DS}. Calculations in this subsection follow \cite{Mahajan:2021nsd}, in which the disk partition function is also considered. 

We consider the Liouville CFT with the central charge $c=\i\,c^{(g)}$, which is parameterized by $b$ as \eqref{eq:liouvillec}. In the classical limit $b\to 0$, we have
\begin{align}
    c\simeq 13+\frac{6}{b^2}\, ,
\end{align}
so the parameter $b$ approximates
\begin{align}\label{eq:approb}
    b^{-2}\simeq\frac{-13+\i\,c^{(g)}}{6}\, .
\end{align}

Let us calculate the partition function in this limit. In the following, we fix the background metric to the unit sphere $\BS^2$:
\begin{align}
    \d s^2=\d\theta^2+\sin^2\theta\,\d\psi^2,\qquad \mathcal{R}=2\, . \label{unitshpere}
\end{align}
In the $b\to0$ limit, the constant solutions to the equation of motion are 
\begin{align}\label{eq:constsolusion}
    2\phi_n=\log\frac{1}{4\mu}+\i\,\pi(2n+1)\, ,\qquad (n\in\mathbb{Z})\, .
\end{align}
The saddle point approximation for a solution associated to $n$ becomes 
\begin{align}
    \int_{\mathcal{C}_n}\d\phi\, e^{-I[\phi]} =-e^{-\i\,\pi(2n+1)/b^2}e\,(4\mu)^{1/b^2+1}\, ,
\end{align}
where $\mathcal{C}_n$ denotes the steepest descent from the saddle point $\phi_n$. Here we have to note that the shapes of contours $\mathcal{C}_n$ and which contours should be summed over depend on the parameter $b$. Considering constant solutions, the original path integral takes the form
\begin{align}\label{eq:constint}
    \int_{-\infty}^{\infty}\d\phi \, \exp\left[-2(b^{-2}+1)\,\phi-4\mu\, e^{2\phi}\right]\, 
\end{align}
with real $b$. To analytically continue to complex $b$, we extend $\phi$ to complex values and regard the integral as along the real axis in the complex plane of $\phi$. However, there are subtleties to analytically continue the parameter such that the resulting integral is certainly well-defined.\footnote{See also \cite{Witten:2010cx,Harlow:2011ny} for details about the method of analytic continuation used here. In particular, appendix C of \cite{Harlow:2011ny} describes the analytic continuation of the gamma function, which takes essentially the same form as the integral considered in this subsection.}

In our limit \eqref{eq:approb}, fortunately, the path integral can be straightforwardly defined because the integrand is convergent in $\phi\to\pm\infty$ when $\text{Re}\left[-2\left(b^{-2}+1\right)\right]>0$. This is a different point from \cite{Mahajan:2021nsd}, where the sign of the real part is reversed. In our case, we can deform the defining contour to the steepest descent $\mathcal{C}_0$ of the $n=0$ saddle point due to the Cauchy's theorem. Therefore the classical approximation of the partition function is 
\begin{align}\label{eq:liouvillesaddle}
    Z_{\text{CFT}}\simeq\int_{\mathcal{C}_0}\d \phi\, e^{-I[\phi]}=-e^{-\i\, \pi/b^2}e\,(4\mu)^{1/b^2+1}\, .
\end{align}
Taking the limit \eqref{eq:approb}, the dominant part in the large $c^{(g)}$ limit is
\begin{align}\label{eq:liouvilleleading}
    Z_{\text{CFT}}\simeq C\,e^{\frac{\pi}{6}c^{(g)}}(4\mu)^{\frac{\i}{6}c^{(g)}}\, ,
\end{align}
where $C$ denotes a constant coefficient  independent of $c^{(g)}$.
We can interpret the factor $e^{\frac{\pi}{6}c^{(g)}}$ as a contribution from the Euclidean part, as we have expected, because the squared norm of the partition function is 
\begin{align}
    \left|Z_{\text{CFT}}\right|^2\simeq\left|C\right|^2e^{\frac{\pi }{3}c^{(g)}}\, ,
\end{align}
which is identical to the gravitational calculation \eqref{vgravity} up to an overall constant. Therefore this result reproduces the relation of the Euclidean dS/CFT \eqref{square} up to an overall constant.

The other factor $(4\mu)^{\frac{\i}{6}c^{(g)}}$ may be interpreted as a contribution from the Lorentzian dS$_3$. Consider the global coordinate of dS$_3$ 
\begin{align}
    \d s^2=-\d T^2+\cosh^2T\,\d\Omega_2^2 \ , \label{globalmetric}
\end{align}
where $\d\Omega_2^2=\d\psi^2+\sin^2\psi\, \d\phi^2$ is the metric of a two-dimensional unit sphere. Introducing a cutoff $T_\infty$, the on-shell action in a region $0<T<T_\infty$ is 
\begin{align}
    I_\textrm{G}=\frac{\Lambda}{4\pi G_N}\int_0^{T_\infty}\d T \int\d\Omega_2\,\sqrt{-g}=\frac{c^{(g)}}{3}\left(T_\infty+\frac{1}{2}\sinh(2T_\infty)\right)\,.
\end{align}
Defining $\epsilon=2e^{-T_\infty}$, the gravity partition function behaves as (see, e.g., \cite{Cotler:2019nbi} as well)
\begin{align}
    Z_{\text{G}}\simeq e^{\frac{\i}{3}c^{(g)}\left(-\log\frac{\epsilon}{2} +\frac{1}{\epsilon^2}\right)}. \label{ZGdiv}
\end{align}
These divergent terms can be thought of as UV divergences in the boundary theory and cancelled by adding local counterterms to the boundary action. While the original partition function is invariant under a Weyl transformation $\delta g^{(0)}=2\delta \sigma g^{(0)}$ and $\delta\epsilon =\delta \sigma \epsilon$ on the boundary, the renormalized partition function is not invariant \cite{Henningson:1998gx};
\begin{align}
    \delta Z_{\text{G}}^{(\text{ren})}=e^{\frac{\i}{3}c^{(g)}\delta \sigma},
\end{align}
because the logarithmic term itself is not invariant and the effective boundary action with divergent terms removed is not. 
This is the same form as $\mu^{\frac{\i}{6}c^{(g)}}$ by setting $\m\propto e^{2\delta\sigma}$. Therefore, the imaginary contribution in \eqref{eq:liouvilleleading} is interpreted as a Weyl anomaly in the boundary theory.

Let us rephrase this conclusion in a slightly different language. With the global coordinates \eqref{globalmetric}, the boundary metric at $T = T_\infty$ may be given by 
$\d\hat s^2 = e^{2 T_\infty} \d s^2 $, where $\d s^2$ is the metric of a unit sphere $\mathbb{S}^2$ given in \eqref{unitshpere}. In order to move from the sphere with radius $e^{T_\infty}$ to the unit sphere, we need to perform the Weyl transformation, where the metric is changed as $\d s^2 \to \d\hat s^2 = e^{\alpha} \d s^2 $. CFT on $\mathbb{S}^2$ should be invariant under the Weyl transformation up to the Weyl anomaly. Due to the Weyl anomaly, the partition function receives a correction as \cite{Friedan:1982is}
\begin{align}
Z^{(\alpha)}_{\text{G}} = e^{I_L} Z^{(\alpha = 0)}_{\text{G}} \, , \qquad
I_L = \frac{c}{48 \pi} \int \d^2 z \left[ \partial \alpha \bar \partial \alpha + \frac{1}{2} \sqrt{g}\, \mathcal{R}\, \alpha \right] \, ,
\end{align}
see \cite{Hikida:2020kil} and appendix \ref{app:DS} for the notation. Therefore, the partition function with $\alpha = 2 T_\infty$ is related to that with $\alpha =0$ up to a phase factor as
\begin{align}
    Z^{(\alpha = 2 T_\infty)}_{\text{G}} = e^{ \frac{\i}{3} c^{(g)} T_\infty } Z^{(\alpha = 0)}_{\text{G}}= e^{ \frac{\i}{3} c^{(g)} \left( - \log \frac{\epsilon}{2} \right)} Z^{(\alpha = 0)}_{\text{G}} \, .
\end{align}
This explains the logarithmic divergence in \eqref{ZGdiv}.

It is easy to extend this calculation to $\text{SU}(N)$ Toda CFT with a central charge $c=\i\,c^{(g)}$, which is dual to a higher-spin gravity described by $\text{SU}(N)$ Chern-Simons gravity. The central charge \eqref{Todac} of $\text{SU}(N)$ Toda CFT approximates 
\begin{align}
    c\simeq N-1+\frac{N(N^2-1)}{b^2}
\end{align}
in $b\to0$. Therefore 
\begin{align}
    b^{-2}\simeq -\frac{1}{N(N+1)}+\frac{\i\,c^{(g)}}{N(N^2-1)}\, .
\end{align}

The action for Toda CFT is rewritten as 
\begin{align}
    I=\frac{1}{2\pi}\int\d^2 w\sqrt{g}\left[\frac{G^{ij}}{2b^2}\,\bar{\partial}\phi_i\partial\phi_j+\frac{\mathcal{R}}{8}\left(b^{-2}+1\right)\sum_{i=1}^{N-1}i(N-i)\phi_i+\mu\sum_{i=1}^{N-1}e^{\phi_i}\right]\, ,
\end{align}
where we have used the relation $\sum_{j}G^{ji}=i(N-i)/2$ for the inverse of Cartan matrix $G^{ij}$ for $\text{SU}(N)$. 
Considering again constant solutions, the partition function approximates
\begin{align*}
    Z_{\text{CFT}}&\simeq \int\d\phi_1\cdots\d\phi_{N-1}\,e^{-\sum_i\left[(b^{-2}+1)i(N-i)\phi_i+4\mu e^{\phi_i}\right]} \\
    &=\prod_{i=1}^{N-1}\int\d\phi_i\, e^{-(b^{-2}+1)i(N-i)\phi_i-4\mu e^{\phi_i}} \, .
\end{align*}
Solutions to the equation of motion at the leading order are 
\begin{align}
    \phi_i=\log\frac{i(N-i)}{4\mu}+\i\,\pi(2n_i+1)\, ,\qquad \left(1\le i\le N-1,\ n_i\in\mathbb{Z}\right) \, .
\end{align}
By the same discussion as in the Liouville CFT, the integral should include only the $n=0$ saddle. Substituting this solution, we obtain the classical approximation 
\begin{align}
    Z_{\text{CFT}}\simeq C\,e^{\frac{\pi}{6}c^{(g)}}\left(4\mu e^{s}\right)^{\frac{\i}{6}c^{(g)}}\, ,
\end{align}
where $C$ is a constant that is independent of $c^{(g)}$ and $s\equiv-\frac{6}{N(N^2-1)}\sum_{i=1}^{N-1}i(N-i)\log \left[i(N-i)\right]$. We can see that the squared norm of the partition function satisfies \eqref{partzero} and the remaining part comes from the Euclidean dS$_3$. When $N=2$, this reproduces the result of Liouville CFT \eqref{eq:liouvilleleading}.  

\subsection{Two-point functions}
Next we focus on two-point functions of the boundary CFT, which can be approximated by geodesic distances in the bulk spacetime. Therefore, we first calculate the length of a geodesic between two points on the boundary of dS$_3$. 

Consider the three-dimensional de Sitter spacetime in Lorentzian signature 
\ba
\d s^2=-\d T^2+\cosh^2 T\, \d\Omega_2^2 \, ,
\ea
where $\d\Omega_2^2=\d\psi^2+\sin^2\psi\, \d\phi^2$ is the metric of a two-dimensional unit sphere.
We are interested in the geodesic distance between two points at the same time: $(T_0,\psi_i,0)$ and $(T_0,\psi_f,0)$. The geodesic distance between the two points is given by
\ba\label{eq:geodesic_distance}
D(\psi_i,\psi_f)=\arccos\left[1-2\sin^2\left(\frac{\psi_f-\psi_i}{2}\right)\,\cosh^2 T_0\right]\, .
\ea
Note that there is no geodesic which connects the two points in the Lorentzian de Sitter spacetime
when $\sin^2\left(\frac{\psi_f-\psi_i}{2}\right)\,\cosh^2 T_0>1$,
where $D(\psi_i,\psi_f)$ gets complex valued.

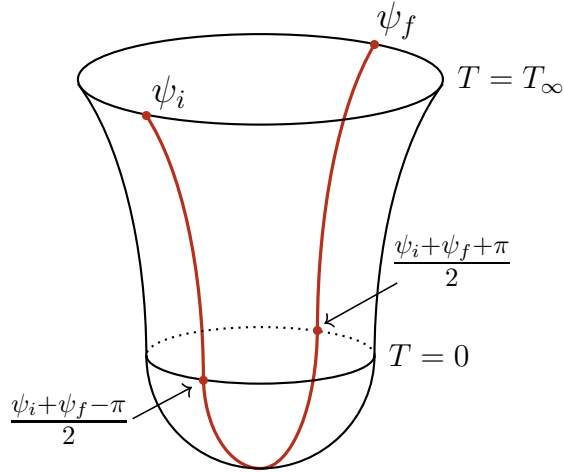
\begin{figure}
  \centering
  \begin{tikzpicture}[thick,scale=0.6]
    \begin{scope}
      \begin{scope}[very thick]
        \draw[yshift=-0.1cm,BrickRed] (2.5,0.87) arc (120:180:2.5 and 7.3);
        \draw[yshift=-0.1cm,BrickRed] (-2.5,-0.7) arc (60:0:2.5 and 6.7);
        \draw[BrickRed] (-1.25,-6.55) arc (180:270:1.25 and 2.05);
        \draw[BrickRed] (1.25,-5.45) arc (0:-90:1.25 and 3.15);
      \end{scope}
      \draw (0,0) ellipse (4 and 1);
      \draw[yshift=-0.1cm] (4,0) arc (120:180:3 and 7);
      \draw[yshift=-0.1cm] (-4,0) arc (60:0:3 and 7);
      \draw (-2.5,-6.1) arc (180:360:2.5);
      \draw (-2.5,-6.1) arc (180:360:2.5 and 0.625);
      \draw[dotted] (2.5,-6.1) arc (0:180:2.5 and 0.625);
      \begin{scope}
        \draw[fill=BrickRed,draw=none,yshift=-0.1cm] (-2.5,-0.7) circle (0.1);
        \draw[fill=BrickRed,draw=none,yshift=-0.1cm] (2.5,0.87) circle (0.1);
        \draw[fill=BrickRed,draw=none,yshift=-0.1cm] (-1.25,-6.55) circle (0.1);
        \draw[fill=BrickRed,draw=none,yshift=-0.1cm] (1.25,-5.45) circle (0.1);
      \end{scope}
      \begin{scope}[very thick]
        \node at (-2,-0.3) {\Large $\psi_i$};
        \node at (3,1.3) {\Large $\psi_f$};
        \node at (4.2,-4) {\Large $\frac{\psi_i+\psi_f+\pi}{2}$};
        \node at (-4.2,-7.5) {\Large $\frac{\psi_i+\psi_f-\pi}{2}$};
        \node at (5.5,0) {\large $T=T_\infty$};
        \node at (3.7,-6.1) {\large $T=0$};
      \end{scope}
      \draw[->,semithick] (3,-4.5)--(1.4,-5.4);
      \draw[->,semithick] (-2.8,-7.5)--(-1.5,-6.8);
    \end{scope}
  \end{tikzpicture}
\caption{The profile of a geodesic which connects two points on the future boundary in the Lorentzian dS$_3$ with an initial Euclidean hemisphere, following from the standard Hartle-Hawking prescription.}
\label{fig:HHgeodesic}
\end{figure}

In the standard dS/CFT, we expect that the dual CFT$_2$ lives on the sphere in the future infinity (see figure \ref{dSCFTfig}), which is located at $T=T_\infty\to \infty$ by introducing the regularization. In this limit we find the geodesic distance:
\ba\label{eq:HHgeodesic}
D(\psi_i,\psi_f)=2\, \i\,T_\infty+\i\,\log\left[ \sin^2\left(\frac{\psi_f-\psi_i}{2}\right) \right] +\pi \, .  \label{geod}
\ea
Even though this does not correspond to a real geodesic in a Lorentzian dS$_3$, we can interpret it as the geodesic length in a geometry obtained by gluing a future half $T>0$ of Lorentzian dS$_3$  with the 
hemisphere $-\frac{\pi}{2}\leq \tau<0$:
\ba
\d s^2 = \d\tau^2+\cos^2 \tau\, \d\Omega_2^2 \, ,
\ea
along the circle $T=\tau=0$ as depicted in figure \ref{fig:HHgeodesic}. The real part $\pi$ in (\ref{geod}) comes from the half circle on the hemisphere, which is identical to the maximal distance of geodesics connecting a pair of boundary points of $\BB^3$, while the imaginary length arises from the time-like geodesics in the Lorentzian dS$_3$.
Indeed, we can show that the total geodesic satisfies the saddle-point condition as follows.
The geodesic distance between $(T_\infty,\psi_i,0)$ and $(T_\infty,\psi_f,0)$ that goes through two points $\tilde{\psi}_i$ and $\tilde{\psi}_f$ on $T=0$ takes the form
\begin{align}\begin{aligned}
    \left|\tilde{\psi}_f-\tilde{\psi}_i\right| 
    + \i\, \log\left[e^{2T_\infty}\cos\left(\psi_i-\tilde{\psi}_i\right)\,\cos\left(\psi_f-\tilde{\psi}_f\right)\right]\,.
\end{aligned}\end{align}
Since this is complex, the extremization condition varying $\tilde{\psi}_i$ and $\tilde{\psi}_f$ splits into those of both real and imaginary parts:
\begin{align}
    \delta\tilde{\psi}_i=\delta\tilde{\psi}_f \, ,\quad \tan\left(\psi_i-\tilde{\psi}_i\right)\,\delta\tilde{\psi}_i+\tan\left(\psi_f-\tilde{\psi}_f\right)\,\delta\tilde{\psi}_f=0 \, ,
\end{align}
which give a relation
\begin{align}
     \psi_i+\psi_f=\tilde{\psi}_i+\tilde{\psi}_f \, .
\end{align}
Assuming the geodesic in $\BB^3$ is maximized to be $\pi$, i.e. satisfying $\tilde{\psi}_f-\tilde{\psi}_i=\pi$, we obtain 
\begin{align}
    \tilde{\psi}_i=\frac{\psi_i+\psi_f-\pi}{2} \, ,\qquad\tilde{\psi}_f=\frac{\psi_i+\psi_f+\pi}{2} \, .
\end{align}
From this condition, one can reproduce our result \eqref{eq:HHgeodesic}.

Now we consider a calculation of two-point functions in the dS$_3/$CFT$_2$. By employing (\ref{geod}),
the geodesic approximation of the gravity dual of the CFT two-point function leads to
\begin{align}
\la\, O(\psi_i)\,O(\psi_f)\,\lb \simeq e^{-\Delta^{(g)} D(\psi_i,\psi_f)}\simeq  \left[\frac{\i\,\ep}{2\sin\left(\frac{\psi_i-\psi_f}{2}\right)}\right]^{2\,\i\,\Delta^{(g)}}
=e^{-\pi\Delta^{(g)}}\cdot\left[\frac{\ep}{2\sin\left(\frac{\psi_i-\psi_f}{2}\right)}\right]^{2\,\i\,\Delta^{(g)}} \, ,
\label{twss}
\end{align}
where the cutoff $\ep$ is defined by $\ep=2e^{-T_\infty}$.
The presence of ``$\i$'' in front of $\ep$ can be understood as an analytical continuation from  AdS as $e^{-\rho_\infty}=2\i \, e^{-T_\infty}$, where $\rho$ is the radial coordinate of global AdS.

Now let us examine if we can reproduce the same result from the CFT side.
The second factor in the right hand side of (\ref{twss}) follows from the standard two-point function in CFT for operators of dimension $\i \, \Delta^{(g)}$. 
Moreover, the first factor $e^{-\pi\Delta^{(g)}}$ can be found from an open Wilson loop 
on the hemisphere $\BB^3$ with two end points on the boundary $\mathbb{S}^2$, evaluated in the light of the CFT dual as follows:
\ba
\frac{Z_\text{CFT}\left[\BB^3,R_j\right]}{Z_\text{CFT}[\BB^3]}=\frac{|\mathcal{S}_{0}^{~j}|}{|\mathcal{S}_0^{~0}|} 
=e^{\frac{\pi}{6}c^{(g)}\left(\s{1-8G_N E_j}-1\right)}  
\simeq e^{-\pi \Delta^{(g)}_j},  \label{corsp}
\ea
assuming $\Delta_j^{(g)}\ll c^{(g)}$, corresponding to the geodesic approximation. In this way, we can reproduce the gravity prediction of the two-point function (\ref{twss}) from the CFT calculation. It is intriguing to note that we can also interpret this two-point function as the standard one in a two-dimensional CFT with an `exotic rule' 
that the UV cutoff is not $\ep$ but is $\i \,\ep$. This might also be interpreted as the radius of the two-sphere, where the CFT lives, is imaginary.

\subsection{Holographic entanglement entropy}
Now we turn to the holographic entanglement entropy. 
By assuming a simple extension of the holographic entanglement entropy in the AdS/CFT \cite{Ryu:2006bv, Ryu:2006ef, Hubeny:2007xt} to the dS/CFT (see 
\cite{Narayan:2015vda,Sato:2015tta,Miyaji:2015yva} for earlier works), we expect that this can be computed from the length of the geodesic. We choose the subsystem $A$ to be the interval 
$\psi_i\leq \psi\leq \psi_f$ at $\phi=0$. Then the holographic entanglement entropy can be found from (\ref{geod}) as
\ba
S_A=\frac{D(\psi_i,\psi_f)}{4G_N}=\i\,\frac{c^{(g)}}{3}\,T_\infty+\i\,\frac{c^{(g)}}{6}\,\log\left[ \sin^2\left(\frac{\psi_i-\psi_f}{2}\right)\right] +\frac{c^{(g)}}{6}\,\pi \, .   \label{heesa}
\ea
We can compare this with the standard formula of entanglement entropy in two-dimensional CFTs
\cite{Holzhey:1994we,Calabrese:2004eu}, which is written as follows in the present setup:
\ba
S_A=\frac{c}{6}\,\log\left[\frac{4\sin^2\left(\frac{\psi_i-\psi_f}{2}\right)}{\ti{\ep}^2}\right] \, .
\ea
Again by setting $c=\i\,c^{(g)}$ and $\ti{\ep}=\i\,\ep = 2\i\,e^{-T_\infty}$, we reproduce the holographic entanglement entropy (\ref{heesa}).

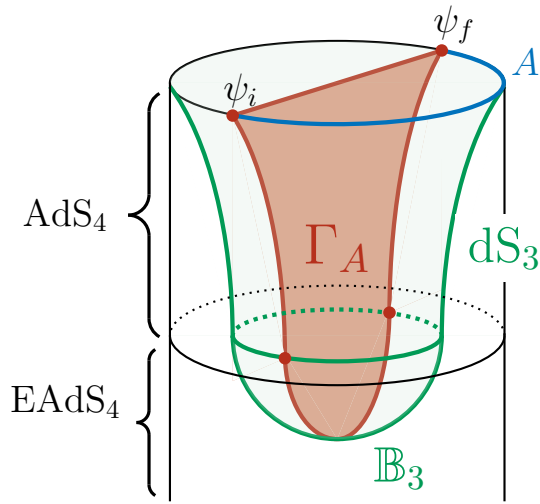
\begin{figure}
  \centering
  \hspace{-2cm}
\begin{tikzpicture}[thick,scale=0.55]
    \begin{scope}
      \begin{scope}[ultra thick] 
        \draw[draw=none,fill=BrickRed!40,yshift=-0.1cm] (-2.5,-0.7)--(-2.5,-7.1)--(2.5,-4.9)--(2.5,0.87)--cycle;
        \draw[yshift=-0.1cm,BrickRed,fill=white] (2.5,0.87) arc (120:180:2.5 and 7.3);
        \draw[yshift=-0.1cm,BrickRed,fill=white] (-2.5,-0.7) arc (60:0:2.5 and 6.7);
        \draw[draw=none,fill=white,yshift=-0.1cm] (-2.51,-0.64) -- (-2.5,-7.1)--(-1.24,-6.55)--cycle;
        \draw[draw=none,fill=white,yshift=-0.1cm] (2.51,0.95)--(2.5,-4.9)--(1.25,-5.45)--cycle;
        \draw[BrickRed,fill=BrickRed!40] (-1.25,-6.55) arc (180:270:1.25 and 2.05);
        \draw[BrickRed,fill=BrickRed!40] (1.25,-5.45) arc (0:-90:1.25 and 3.15);
        \draw[draw=none,fill=BrickRed!40] (-1.25,-6.55)--(0,-8.6)--(1.25,-5.45)--cycle;
        \draw[BrickRed,yshift=-0.1cm] (-2.5,-0.7) -- (2.5,0.87);
      \end{scope}
      \draw (0,0) ellipse (4 and 1);
      \begin{scope}[ultra thick] 
        \draw[draw=none,fill=ForestGreen!20,opacity=0.2] (-4,0)--(-4,-6.1)--(4,-6.1)--(4,0)--cycle;
        \draw[ForestGreen] (-2.5,-6.1) arc (180:360:2.5);
        \draw[draw=none,fill=ForestGreen!20,opacity=0.2] (-2.5,-6.1) arc (180:360:2.5);
        \draw[ForestGreen] (-2.5,-6.1) arc (180:360:2.5 and 0.625);
        \draw[dotted,ForestGreen] (2.5,-6.1) arc (0:180:2.5 and 0.625);
        \draw[draw=none,fill=white] (-4,0)--(-4,-6.1)--(-2.5,-6.1)--cycle;
        \draw[draw=none,fill=white] (4,0)--(4,-6.1)--(2.5,-6.1)--cycle;
        \draw[yshift=-0.1cm,ForestGreen,fill=white] (-4,0) arc (60:0:3 and 7);
        \draw[yshift=-0.1cm,ForestGreen,fill=white] (4,0) arc (120:180:3 and 7);
        \draw[draw=none,fill=ForestGreen!20,opacity=0.2] (4,0) arc (0:180:4 and 1);
      \end{scope}
      \draw (4,0) -- (4,-10.1);
      \draw (-4,0) -- (-4,-10.1);
      \draw (-4,-6.1) arc (180:360:4 and 1.2);
      \draw[dotted] (4,-6.1) arc (0:180:4 and 1.2);
      \begin{scope}[ultra thick]
        \draw[RoyalBlue,yshift=-0.1cm] (2.5,0.87) arc (51:-130:4 and 1);
      \end{scope}
      \begin{scope} 
        \draw[fill=BrickRed,draw=none,yshift=-0.1cm] (-2.5,-0.7) circle (0.15);
        \draw[fill=BrickRed,draw=none,yshift=-0.1cm] (2.5,0.87) circle (0.15);
        \draw[fill=BrickRed,draw=none,yshift=-0.1cm] (-1.25,-6.55) circle (0.15);
        \draw[fill=BrickRed,draw=none,yshift=-0.1cm] (1.25,-5.45) circle (0.15);
      \end{scope}
      
      \begin{scope}[ultra thick] 
        \node at (-2.3,-0.2) {\Large $\psi_i$};
        \node at (2.8,1.4) {\Large $\psi_f$};
         \node[RoyalBlue] at (4.5,0.5) {\Large $A$};
         \node[ForestGreen,scale=1.3] at (1.5,-9.3) {\Large $\BB_3$};
         \node[ForestGreen,scale=1.3, fill=white] at (4,-4) {\Large dS$_3$};
         \node[BrickRed,scale=1.5] at (0,-4) {\Large $\Gamma_A$};
      \end{scope}
      \begin{scope}[shift={(-4.5,-3.2)},very thick,scale=0.5]
          \draw (-1,0) arc (270:360:1);
          \draw (0,1)--(0,5);
          \draw (0,5) arc (180:120:1);
          \draw (-1,0) arc (90:0:1);
          \draw (0,-1) -- (0,-5);
          \draw (0,-5) arc (180:240:1);
      \end{scope}
      \begin{scope}[shift={(-4.5,-8.2)},very thick,scale=0.3]
          \draw (-1,0) arc (270:360:1);
          \draw (0,1)--(0,5);
          \draw (0,5) arc (180:120:1);
          \draw (-1,0) arc (90:0:1);
          \draw (0,-1) -- (0,-5);
          \draw (0,-5) arc (180:240:1);
      \end{scope}
      \node at (-6.5,-3.2) {\Large AdS$_4$};
      \node at (-6.5,-7.7) {\Large EAdS$_4$};
    \end{scope}
  \end{tikzpicture}
\caption{A sketch of calculation of holographic entanglement entropy in the codimension-two holography.
The green surface describes the dS$_3$ brane, where we impose the Neumann boundary condition and we expect the presence of three-dimensional gravity. The subsystem $A$ is defined on the $\mathbb{S}^2$ on the boundary of the dS$_3$. The red surface describes the extremal surface $\Gamma_A$ whose area leads to the holographic entanglement entropy. The four-dimensional ambient spacetime is AdS$_4$.}
\label{braneworldfig}
\end{figure}

We can also confirm the result (\ref{heesa})
by applying the codimension-two holography (or wedge holography) introduced in \cite{Akal:2020wfl}. Refer to figure \ref{braneworldfig} for a sketch of our setup.
This has an advantage that we can apply the AdS/CFT, which is more established, rather than the dS/CFT. 
For our purpose, we start with the global AdS$_4$ spacetime described by the de Sitter slices:
\ba
    \d s^2 = \d\eta^2+\sinh^2\eta\,(-\d T^2+\cosh^2 T\, \d\Omega_2^2) \, .
\ea
We limit the value of $\eta$ to $0\leq \eta\leq \eta_*$ and impose the Neumann boundary condition on the boundary $\eta=\eta_*$. According to the idea of brane-world holography \cite{Karch:2000ct}, this gravity background is dual to a quantum gravity on dS$_3$ defined by $\eta=\eta_*$. As before, we also cut along $T=0$ and glue the geometry $T>0$ with the hemisphere obtained by the Wick rotation $T\to \tau = \i\,T$.  

The codimension-two holography argues that this four-dimensional geometry is dual to a CFT on the two-dimensional sphere at $T=T_{\infty}\to\infty$. In this setup, the entanglement entropy $S_A$ of the two-dimensional CFT is computed by the area of the extremal surface such that it ends on the boundary $\de A$ of the subsystem $A$. This extremal surface $\Gamma_A$ is simply given by the two-dimensional surface defined by the product of the geodesic in dS$_3$ (i.e. the one in figure \ref{fig:HHgeodesic}) times the interval of the $\eta$ direction 
given by $[0,\eta_*]$. 

The Newton constant in the effective three-dimensional gravity can be found from that in four dimensions
via the dimensional reduction as follows:
\ba
\frac{1}{G_N}=\frac{1}{G^{(4)}_N}\int^{\eta_*}_{0}\d \eta\,\sinh\eta=\frac{2\cosh^2\frac{\eta_*}{2}}{G^{(4)}_N} \, .
\label{reduceKK}
\ea
The holographic entanglement entropy in the codimension-two holography reads 
\begin{align}
S_A=\frac{\mbox{Area}(\Gamma_A)}{4G^{(4)}_N}=\frac{1}{4G^{(4)}_N}\int^{\eta_*}_{0} \d\eta\,\sinh\eta\cdot  D(\psi_i,\psi_f) \, .
\end{align}
This indeed reproduces (\ref{heesa}) by noting the relation (\ref{reduceKK}).

\section{Relation to Gaberdiel-Gopakumar duality}
\label{sec:GG}

Up to now, we have examined the duality between the classical limit of (higher-spin) gravity on $\mathbb{S}^3$ and SU(2) (or SU$(N)$) WZW model at the critical level limit. For the purpose, we do not need details of duality, namely, how $M_\text{CFT}$ is realized in \eqref{dualitytwo} (or \eqref{dualityN}). In this section, we provide a concrete realization as an analytic continuation of Gaberdiel-Gopakumar duality \cite{Gaberdiel:2010pz}. Referring to the realization, we discuss the spectrum of the WZW model with non-integer $k$ and comment on quantum corrections in the succeeding sections.

\subsection{Original Gaberdiel-Gopakumar duality}

Before going into the details of our proposal, we would like to review the original Gaberdiel-Gopakumar duality \cite{Gaberdiel:2010pz}. The duality is between a higher-spin gravity in three dimensions and the two dimensional $W_N$-minimal model described by the coset \eqref{cosetc}.
The higher-spin gravity is given by Prokushkin-Vasiliev theory \cite{Prokushkin:1998bq} with higher-spin gauge fields with spin $s=2,3,\ldots$ and two complex massive scalars. The proposal is that the classical limit of the higher-spin gravity corresponds to the large $N,k$ limit of the coset CFT \eqref{cosetc} but the 't Hoof parameter
\begin{align}
    \lambda = \frac{N}{k + N} \label{thooft}
\end{align}
fixed finite. It was confirmed by the match of symmetry, partition function, correlation function, and so on, see \cite{Gaberdiel:2012uj} for a review.

The higher-spin gravity constructed by \cite{Prokushkin:1998bq} is defined on three-dimensional AdS space with a negative cosmological constant, but it is not difficult to move to dS space with a positive cosmological constant by a simple analytic continuation.
Gravity theory on AdS$_3$ is topological and it can be described by a Chern-Simons theory \cite{Achucarro:1986uwr,Witten:1988hc}.
The action of the Chern-Simons gauge theory is given by
\begin{align}
	I_\text{CSG}= I_\text{CS} [A] - I_\text{CS} \left[\bar A\right] \, , \quad
	I_\text{CS}[A]  = \frac{k_\text{CS}}{4 \pi} \int_{\mathcal{M}} \text{Tr} \left( A \wedge \d A + \frac{2}{3}\, A \wedge A \wedge A \right) \, . \label{CSactionAdS}
\end{align}
The gauge fields $A, \bar A$ take values in $\mathfrak{sl}(2)$ Lie algebra and the Chern-Simons level $k_\text{CS}$ is related to the Newton constant $G_N$ and the AdS radius $L_\text{AdS}$ as $k_\text{CS} = L_\text{AdS}/(8 G_N \epsilon_N)$, see \eqref{hsk2} for the $\mathbb{S}^3$ case.
With this description, we can easily construct a higher-spin gravity by replacing $\mathfrak{sl}(2)$ with a higher rank Lie algebra $\mathfrak{g}$. 
With the choice of $\mathfrak{g} = \mathfrak{sl}(N)$, we obtain a higher-spin gravity on AdS$_3$ with gauge fields of spin $s=2,3,\ldots,N$. Note that an analytic continuation of the level $k_\text{CS} \to \kappa = - \i \, k_\text{CS}$ yields a higher-spin gravity on dS$_3$ described as in \eqref{CSaction0} with \eqref{CSaction}.%
\footnote{The action \eqref{CSaction0} with \eqref{CSaction} is actually for a higher-spin gravity on $\mathbb{S}^3$, thus we need further analytic continuation for the time direction.}
The gauge algebra of the Prokushkin-Vasiliev theory is an infinite dimensional higher-spin algebra denoted by $\mathfrak{hs}[\lambda]$. One way to define $\mathfrak{hs}[\lambda]$ is extending $\mathfrak{sl}(N)$ such that $\mathfrak{hs}[\lambda]$ reduces to $\mathfrak{sl}(N)$ at $\lambda = N$ by dividing its ideal. The Prokushkin-Vasiliev theory includes two complex scalars with mass
\begin{align}
 L_\text{AdS}^2\, m^2 = - 1 + \lambda^2
\end{align}
as well.
The parameter $\lambda$ is the same as the parameter for the gauge algebra $\mathfrak{hs}[\lambda]$, and it is identified with the 't Hooft parameter introduced in the coset CFT \eqref{cosetc} as \eqref{thooft}. 
Here we set
\begin{align}
0 \leq \lambda \leq 1 \, , \label{lambdarange}
\end{align}
which is required from the dual CFT.
For the range of parameter $\lambda$, 
we can assign different boundary conditions to these two complex scalars. The conformal dimensions of the dual operators are then given by
\begin{align}
\Delta_\pm =1 \pm \lambda \label{dualcd}
\end{align}
with the choice of boundary conditions.

We would like to provide some supporting arguments for the proposal. Firstly, we examine the match of symmetry. The asymptotic symmetry of higher-spin gravity on AdS$_3$ has been analyzed in 
\cite{Henneaux:2010xg,Campoleoni:2010zq,Gaberdiel:2011wb,Campoleoni:2011hg}. For pure gravity on AdS$_3$, the asymptotic symmetry is well-known to be the Virasoro algebra with the central charge \cite{Brown:1986nw}
\begin{align}
    c = 6\, k_\text{CS} = \frac{3\, L_\text{AdS}}{4\, G_N \epsilon_N} \, , \label{BHc}
\end{align}
see \eqref{centNa} for the $\mathbb{S}^3$ case.
For higher-spin gravity with $\mathfrak{sl}(N)$ Lie algebra, the asymptotic symmetry is given by $W_N$-algebra with the same central charge \eqref{BHc}. In the same way, the asymptotic symmetry of the Prokushkin-Vasiliev theory is given by an infinite-dimensional higher-spin algebra called as $W_\infty [\lambda]$-algebra again with the same central charge \eqref{BHc}.
The algebra $W_\infty[\lambda]$ truncates at $\lambda =  L$ $(L =2,3,\ldots)$ to $W_L$-algebra with spin $s=2,3,\ldots,L$. The dual coset \eqref{cosetc} has $W_N$-symmetry and is assigned to have the same central charge. 
Since $\lambda$ is in the range of \eqref{lambdarange}, $W_\infty[\lambda]$ and $W_N$ does not look like the same algebra. However, it was shown in \cite{Gaberdiel:2012ku} that the algebra $W_\infty[\lambda]$ has the triality relation
\begin{align}
W_\infty[\tfrac{N}{N+k}] \simeq W_\infty[- \tfrac{N}{N+k+1}]  \simeq W_\infty[N] \ , \label{triality}
\end{align}
with the same central charge $c$.
With the help of the triality relation, we can show the match of symmetry algebras for the duality.
It might be useful to note that this triality is equivalent to the following two duality relations:
\ba
&& (a)\qquad (k',N',\lambda')= \left(-2N-k-1,N,-\frac{N}{N+k+1}\right) \, ,\\  \label{traa}
&& (b)\qquad (k',N',\lambda')=\left(\frac{1-N}{N+k},\frac{N}{N+k},N\right) \, .\label{trab}
\ea
These two relations generate totally six pairs of $(k,N)$ with the same central charge:
\begin{align}
        \begin{tikzpicture}[>=stealth, scale=0.8]
        \node (one) {$(k,N)$};
        \node (two) [below=2cm of one] {$\displaystyle \left(\frac{1-N}{N+k}, \frac{N}{N+k}\right)$};
        \node (three) [right=4cm of one] {$(-2N-k-1, N)$};
        \node (four) [below=2cm of two] {$\displaystyle \left(-\frac{2N+k+1}{N+k}, \frac{N}{N+k}\right)$};
        \node (five) [below=2cm of three] {$\displaystyle \left(\frac{N-1}{N+k+1}, -\frac{N}{N+k+1}\right)$};
        \node (six) [below=2cm of five] {$\displaystyle \left(-\frac{k}{N+k+1}, -\frac{N}{N+k+1}\right)$};
        \draw[<->] (one) to node [above] {$a$} (three);
        \draw[<->] (one) to node[left] {$b$} (two);
        \draw[<->] (two) to node[left] {$a$} (four);
        \draw[<->] (three) to node [right] {$b$} (five);
        \draw[<->] (five) to node [right] {$a$} (six);
        \draw[<->] (six) to node [below] {$b$} (four);
    \end{tikzpicture}
\end{align}

As another evidence for the duality, we examine the spectrum of the coset CFT \eqref{cosetc} and check that it reproduces \eqref{dualcd}.
The primary state of the coset CFT can be obtained from the states in the WZW models constructing the coset, see \cite{DiFrancesco:1997nk} for details.
Since the coset \eqref{cosetc} consists of three SU$(2)$ WZW models, a primary state is labeled by three Young diagrams $(\mu , \xi ; \nu)$.
In general, all possible Young diagrams are not allowed for coset CFT, and selection rules and field identifications have to be taken care of.
In the current case, $\xi$ is uniquely fixed by $\mu ,\nu$, so a primary state can be labeled by two Young diagrams as $(\mu ; \nu)$.
The conformal weight of a primary state $(\mu ; \nu)$ is given by
\begin{align}
    h_{(\mu;\nu)} = n + h^{(k)}_\mu + h^{(1)}_\xi - h^{(k+1)}_\nu \, , \qquad
    h^{(k)}_\mu = \frac{C_2 (\mu)}{k +N} \, , \qquad C_2 = \frac{1}{2}\,(\mu , \mu + 2 \rho) \, , \label{generich}
\end{align}
see \eqref{hsh}. The non-negative integer $n$ is determined by how the numerator representation is embedded in the denominator one, see \cite{DiFrancesco:1997nk} for details. The two bulk complex scalars are dual to the CFT operators labeled with $(\square ; 0)$,  $(0 ;\square)$ or their complex conjugates. The conformal weights can be computed as
\begin{align}
 h_{ (\square ; 0) } = \frac{(N-1)}{2 N} \left( 1 + \frac{N+1}{N+k}\right) \, , \qquad 
 h _ {(0 ; \square ) } = \frac{(N-1)}{2 N} \left( 1 - \frac{N+1}{N+k + 1 }\right) \, .  \label{cd}
\end{align}
Taking the 't Hoof limit, they become
\begin{align}
h _ { (\square ; 0) } = \frac{1}{2}\, (1 + \lambda) \, , \qquad h _ {  ( 0 ; \square ) } = \frac{1}{2}\, (1 -\lambda) \, .
\end{align}
Thus we have shown $\Delta_+ =2 h _ { (\square ; 0) }$ and $\Delta_- =2 h _ { (0;\square ) }$ in the limit, which is consistent with the duality.
Generic states of the coset CFT are supposed to correspond to the bound states of scalar fields in the higher-spin theory.

\subsection{Our duality from higher-spin \texorpdfstring{AdS$_3$}{AdS3} holography}

In \cite{Anninos:2011ui}, a dS$_4$/CFT$_3$ correspondence was proposed as an ``analytic continuation'' of higher-spin AdS$_4$ holography of \cite{Klebanov:2002ja}. Thus, it is natural to expect that a similar dS$_3$/CFT$_2$ correspondence can be proposed by taking an ``analytic continuation'' of the Gaberdiel-Gopakumar duality reviewed in the previous subsection, see \cite{Ouyang:2011fs} for an early attempt. To interpret our duality proposed in subsection \ref{sec:dSCFT} as an analytically continued duality, it is convenient to utilize the Gaberdiel-Gopakumar duality viewed in a different way as in \cite{Castro:2011iw,Gaberdiel:2011wb,Perlmutter:2012ds}. In this subsection, we introduce the different viewpoint, then it is almost straightforward to see the relation to our proposal.

As explained in the previous subsection, the match of symmetry for the original Gaberdiel-Gopakumar duality is a consequence of the triality relation of $W_\infty[\lambda]$-algebra \eqref{triality}. Since the symmetry of $\text{SL}(N) \times \text{SL}(N)$ Chern-Simons theory is the $W_N$-algebra as mentioned above, it is natural to conjecture a holographic duality between the $W_N$-minimal model described by the coset \eqref{cosetc} and the Chern-Simons theory.
However, in order to have a duality with classical higher-spin gravity, we need to realize a large central charge in the dual CFT.
From the formula of central charge \eqref{centwm}, we can see that the central charge does not exceed $N-1$ 
for the coset CFT \eqref{cosetc} with positive integer $k$.
In \cite{Castro:2011iw}, a large central charge is realized by performing an analytic continuation of the level $k$ as%
\footnote{We can similarly take the limit of $k \to - N -1$ to obtain a large central charge limit since the expression \eqref{centwm} diverge both for $k \to -N$ and $k \to -N -1$. For $k \to -N$, SU$(N)_k$ in the numerator of \eqref{cosetc} gives dominant contributions. On the other hand, for $k \to - N -1$, SU$(N)_{k+1}$ in the denominator of \eqref{cosetc} gives dominant contributions.}
\begin{align}
k = - N - \frac{N (N^2 -1)}{c}  + \mathcal{O} ( c^ {-2} ) \ , \label{limit}
\end{align}
with keeping $N$ fixed. One may feel uncomfortable to take a non-integer $k$ in the coset CFT \eqref{cosetc}. However, this analytic continuation can be justified at the level of algebra \cite{Gaberdiel:2011wb}.
Namely, we just need to consider a larger algebra $W_\infty[\lambda]$ with large $c$ and then to set $\lambda =N$ with dividing an ideal formed.
If we furthermore set $k$ to an integer, then an additional ideal forms. Dividing the additional ideal, a minimal model with respect to the $W_N$-algebra is constructed, see, e.g., \cite{Prochazka:2015deb}.

To figure out what the dual higher-spin gravity looks like at the corresponding limit, we examine the spectrum of the CFT at \eqref{limit}. The conformal dimensions of the two fundamental states were computed in \eqref{cd}. Taking the limit of \eqref{limit}, we find
\begin{align}
 h _{(\square ; 0)} = - \frac{c}{2N^2} + \mathcal{O} (c^0) \, , \qquad  h _{(0 ; \square  )} = \frac{1 - N}{2 } + \mathcal{O} (c^{-1}) \, .  \label{cdl}
\end{align}
The first one is proportional to the central charge, and the corresponding state is usually dual to a classical configuration. Actually, the conformal weights of generic states become
\begin{align}
 h_{(\mu ; \nu)} = - \frac{  c \, C_2 (\mu)}{N (N^2 -1)}  + \mathcal{O} (c^0) \, , 
\end{align}
and the spectrum has been reproduced by conical defect geometry (or precisely speaking conical surplus geometry) constructed in \cite{Castro:2011iw}. The match of higher-spin charges has been shown as well. 
Note that the leading contribution comes from $\SU(N)_k$ in the coset \eqref{cosetc} since it gives the leading contribution due to the factor $1/(k+N)$. The second one is of order $\mathcal{O} (c^0)$, so it should be regarded as a perturbative complex scalar (combined with the scalar dual to CFT state $(0; \bar \square)$).
These conformal dimensions are negative, and it corresponds to the fact that the CFT at the limit \eqref{limit} is non-unitary.
Generic states with label $(\mu;\nu)$ are identified with bound states with conical defects and perturbative scalar fields \cite{Castro:2011iw,Perlmutter:2012ds}. 
In summary, the coset CFT \eqref{cosetc} at the limit \eqref{limit} is conjectured to be dual to the $\SL(N) \times \SL(N)$ Chern-Simons gauge theory coupled with a perturbative complex scalar with dual conformal dimension $\Delta = 1 -N$.
An important point here is that we need to sum over conical defect geometries as non-perturbative contributions in the higher-spin theory.

We would like to construct an dS$_3$/CFT$_2$ by performing an analytic continuation of this viewpoint of Gaberdiel-Gopakumar duality and see the relation to our proposal summarized in subsection \ref{sec:dSCFT}. 
As mentioned above, the Chern-Simons description of higher-spin gravity on dS$_3$ is given by replacing $k_\text{CS}$ with $ \i \, \kappa$ satisfying $\kappa \in \mathbb{R}$. This implies that gauge sector is organized by $\text{SL}(N) \times \text{SL}(N)$ Chern-Simons gauge theory 
and the asymptotic symmetry is given by $W_N$-algebra but with a pure imaginary central charge (\ref{centNa}) as analyzed in \cite{Ouyang:2011fs}.
The symmetry can be realized by a coset CFT \eqref{cosetc} with imaginary central charge. Setting $c = \i \, c^{(g)}$ with $c^{(g)} \in \mathbb{R}$, the limit \eqref{limit} becomes \eqref{limit0}.
As in the case of AdS$_3$/CFT$_2$, the SU$(N)_k$ sector in the coset \eqref{cosetc} dominates at the limit \eqref{limit0}. In other words, we have
\begin{align}
    M_\text{CFT} = \widehat{\SU(N)}_1 /\widehat{\SU(N)}_{k+1} \label{extra}
\end{align}
in this case. At large central charge $c^{(g)}$, the dominant contribution to the conformal weight becomes
\begin{align}
 h _\mu =  - \i \, \frac{  c^{(g)}  C_2 (\mu)}{N (N^2 -1)}  = \i \, h^{(g)}_\mu \ ,\label{hrel}
\end{align}
under the analytic continuation of central charge.
Note that the conformal weight is also pure imaginary.
As shown in section \ref{sec:hs}, the state labeled by the Young diagram $\mu$ corresponds to a conical defect (surplus) solution associated with the same Young diagram. In the case of AdS$_3$/CFT$_2$, a complex scalar field gives contributions to the higher-spin sector in the next order of $1/c^{(g)}$.
We expect that the situation is similar even in the current dS$_3$/CFT$_2$ case. This will be discussed in section \ref{sec:qc} below.

\section{CFT and black hole spectra}
\label{sec:spectrum}

As in \eqref{dualitytwo}, we have claimed that SU$(2)$ WZW model at the critical level limit $k \to -2$ corresponds to a pure gravity on dS$_3$ at the classical limit.  The primary state is labeled by the representation $R_j$, which is related to the energy $E_j$ of the dual object as
\begin{align}
L\, E_j  = - \frac{c^{(g)}}{3} j (j+1) \, , \label{enegyrel2}
\end{align}
see \eqref{hjsu2} and \eqref{enegyrel}. The dual object is expected to be a black hole solution with the metric
\begin{align}
    \d s^2=L^2\left[(1-8G_N E_j -r^2)\,\d\tau^2+\frac{\d r^2}{1-8G_N E_j -r^2}+r^2\, \d \phi^2\right]  \label{dSBH2}
\end{align}
as in \eqref{dSBH}. As argued in section \ref{sec:gravity}, the solution also corresponds to a conical geometry with deficit angle 
\begin{align}
\delta_j = 2\pi\, \left(1 - \sqrt{1 - 8 G_N E_j} \right) \, . \label{dangle}
\end{align}
However, we have not so far specified the value of $j$ in the CFT or possible black hole energy $E_j$. For a positive integer $k$, the unitary representation is given by $j=0,1/2,1,\ldots,k/2$. However, now $k$ is a complex number, and the unitarity of CFT is violated. Therefore, in principle, other choices are allowed. In this section, we first consider the SU$(2)$ case and examine the possible spectrum and its dual interpretation. We then mention the SU$(N)$ case with generic $N$ and the relation to the Gaberdiel-Gopakumar duality.

\begin{table}
 \centering
  \begin{tabular}{|c|l|l|l|}
   \hline
   ~ & The value of $j$ &Possible rep. & Black hole interpretation \\
   \hline    \hline
   (1) & $j=0$ &Identity rep. & 3-sphere $(\mathbb{S}^3)$\\
   \hline  
   (2) & $j=1/2,1,\ldots$ &Degenerate rep. &  Conical surplus ($\delta < 0$), discrete spectrum \\
   \hline  
  (3) & $j > 0$ &Discrete rep. & Conical surplus ($\delta < 0$), continuous spectrum \\
   \hline   
  (4) & $-1/2 < j < 0$ &Complementary rep. &  Conical defect ($\delta > 0$), continuous spectrum \\
   \hline  
  (5) & $j = -1/2 + \i \, \mathbb{R}_+$ &Continuous rep.&  No clear geometrical interpretation\\
   \hline  
\end{tabular}
 \caption{The possible value of $j$ for the SU$(2)$ representation and its dual interpretation of black hole spectrum.}
 \label{table:spectrum}
\end{table}

We wish to use the representations of the $\SU(2)_k$ WZW model such that the dual objects have a geometrical interpretation. A necessary condition may be the reality of the energy, i.e., $E_j \in \mathbb{R}$, and we will assume the reality condition in the following. We may classify the range of $j$ as in the Table~\ref{table:spectrum}. Here we have restricted $j \geq -1/2$ or $j = -1/2 + \i \, \mathbb{R}_+$ using the symmetry under $j \to - j - 1$. The case (1) is with $j=0$, which corresponds to the identity representation. This leads to the vacuum state or 3-sphere in the dual gravity theory. The case $(2)$ is with $j=1/2,1,3/2,\ldots$, which corresponds to the naive analytic continuation of the unitary representation of SU$(2)$. Note that the upper bound is removed for generic $k$. The dual geometry is given by a conical surplus with deficit angle $\delta < 0$. The black hole mass only takes discrete values satisfying \eqref{enegyrel2}.
The case (3) is with $j > 0$, which may correspond to the discrete series, and the energy $E_j$ takes a negative real value. The black hole geometry \eqref{dSBH} is almost the same as the case (2) and the only difference is now the spectrum is continuous. The case (4) is with $-1/2 < j < 0$, which would correspond to the complementary or discrete series, and the energy takes
\begin{align}
0 < 8 G_N E_j < 1 \, .
\end{align}
The dual object can be regarded as a conical geometry with deficit angle $\delta_j > 0 $. Since the deficit angle is positive, the solution is physical. After the analytic continuation, it is expected to correspond to a physical black hole solution.
The case (5) is with $j = - 1/2 + \i \,  \mathbb{R}_+$, which corresponds to the continuous series. This choice is natural for Liouville description of the coset \eqref{cosetc}. However, the black hole energy takes
\begin{align}
8 G_N E_j \geq 1
\end{align}
and the signature of metric becomes $(-,-,+)$. Thus, the geometrical interpretation of the solution \eqref{dSBH2} becomes obscure.

In the previous section, we realized our duality as an analytic continuation of the Gaberdiel-Gopakumar duality. This actually suggests the cases (1) and (2), namely, the natural analytic continuation of the integrable representation of the affine SU(2) without an upper bound of $j$. In fact, we characterized the dual geometry so as to satisfy the condition of trivial holonomies \eqref{trivilahol}, and we found that the geometry can be classified by an integer value for $N=2$. This implies that we have to sum over the black hole solutions \eqref{dSBH} with discrete values of the energy $E_j$ satisfying the condition \eqref{E2j} for $j=0,1/2,1,\ldots$.
For a generic $N$, the dual geometry is labeled by a Young diagram $\mu$, which corresponds to the integrable representation of the affine SU$(N)$ without the upper bound of $|\mu|$.
Note that, however, our proposal is actually more generic as in \eqref{dualitytwo} or \eqref{dualityN}.
So far we have only considered the leading order in $1/c^{(g)}$ and the additional CFT $M_\text{CFT}$ is not relevant to this order. Thus, we do not need to rely on the Gaberdiel-Gopakumar duality, and a more generic spectrum of the SU$(N)_k$ WZW model could be adopted at least at this order. In the language of the dual higher-spin gravity, it is our choice which kind of geometry is summed over. Namely, we can use gravity solutions which do not satisfy the trivial holonomy condition as well.

\section{Comments on quantum corrections}
\label{sec:qc}

As evidence for our proposal, we have computed gravity partition functions both from the SU$(N)$ WZW model at the leading order in $1/c^{(g)}$ and the classical limit of (higher-spin) gravity and find perfect matches. It is natural to wonder what happens if we go beyond the leading order in $1/c^{(g)}$ or the classical limit. It is actually not an easy task. In the proposal of \eqref{dualityN}, only the sector of the SU$(N)_k$ WZW model matters at the leading order but $M_\text{CFT}$ starts to give contributions at the next order. In the same way, the partition functions can be obtained purely from the (higher-spin) gravity in the classical limit, but in principle there would be contributions from other perturbative matters at the next leading order. This is indeed what happens when our duality is realized as an analytic continuation of the Gaberdiel-Gopakumar duality as explained in section \ref{sec:GG}. In this section, we explain this fact more quantitatively. We focus only on the vacuum partition function, but the situation of other partition functions is qualitatively similar.

As usual we begin with the simplest case with $N=2$.
In section \ref{sec:gravity}, we computed the gravity partition function from the SU$(2)$ WZW model at the level
$k \to -2 + 6 \, \i \, /c^{(g)} $ and considered the large $c^{(g)}$ limit as in \eqref{cricialN2}. For a while, let us forget about the other sector $M_\text{CFT}$ introduced in \eqref{dualitytwo}. The partition function is actually obtained with finite $k$ from the dual CFT as
\begin{align}
Z_\text{CFT} = \left|\Smat{0}{0}\right|^2 = \left|\s{\f{2}{k+2}}\,\sin\left(\frac{\pi}{k+2}\right)\right|^2 \ ,\label{vexact}
\end{align}
see \eqref{SU2S} and \eqref{S00}.  The partition function may be expanded as
\begin{align}
Z_\text{CFT} = e^{- I_\text{CFT}} \, , \qquad I_\text{CFT} = I_\text{CFT}^{(0)} + I_\text{CFT}^{(1)} \log c^{(g)} + \cdots \, .
\end{align}
The leading order expression $ I_\text{CFT}^{(0)} $ can be read off from \eqref{vpartition}. The next leading order can be also obtained from  \eqref{vpartition} as 
\begin{align}
I_\text{CFT}^{(1)} = -1 \, .
\end{align}
We denote the partition function computed from gravity theory by $Z_\text{G}$ and expand it as
\begin{align}
Z_\text{G} = e^{- I_\text{G}} \, , \qquad I_\text{G} = I_\text{G}^{(0)} + I_\text{G}^{(1)} \log c^{(g)} + \cdots \, .
\end{align}
The leading order contribution $I_\text{G}^{(0)}$ was obtained in \eqref{vgravity} and it was shown to agree with $I_\text{CFT}^{(0)}$.
The next order contribution $I_\text{G}^{(1)}$ may be found in \cite{Anninos:2020hfj}
as 
\begin{align}
I_\text{G}^{(1)} = 3 \, . \label{IG1}
\end{align}
Therefore, we do not find match at this order.
However, this is not a contradiction since we have neglected $M_\text{CFT}$ in the CFT side and perturbative matter fields in the gravity side.

Before discussing the discrepancy, let us comment on the Chern-Simons description of pure gravity on dS$_3$ \cite{Witten:1988hc}.
The papers \cite{Castro:2011xb} and \cite{Anninos:2020hfj} argue that the correct 
de Sitter partition function computed from the $\SU(2)\times \SU(2)$ Chern-Simons gauge theories 
is  given by 
\ba
Z_\text{CS}\left[\mathbb{S}^3\right]=\left|\s{\f{2}{k+2}}\,\sin\left(\frac{\pi}{k+2}\right)\right|^2 e^{2 \, \i \, \pi k} \, .
\ea
Note that the new factor $e^{2 \, \i \, \pi k}$ comes from the on-shell Chern-Simons action for the classical background of $\mathbb{S}^3$ and this factor is indeed trivial for integer $k$.
In the large-$k$ limit, this leads to
\begin{align}
Z_\text{CS}\left[\mathbb{S}^3\right]\simeq \frac{1}{(c^{(g)})^3}\,e^{\frac{\pi}{3}\,c^{(g)} } \, ,
\end{align}
where $c^{(g)}=6 \, \i \, k$. From this expression, we can read off $I_\text{G}^{(0)}$ as in \eqref{vgravity} and $I_\text{G}^{(1)}$ as in \eqref{IG1}.

Let us come back to the discrepancy. As argued above, we have to include $M_\text{CFT}$ if we would like to examine the next order effects in $1/c^{(g)}$ in the CFT side. As a concrete example, we regard SU$(2)_k$ WZW model as a part of coset CFT \eqref{cosetc} with $N=2$ describing (an analytic continuation of) the Virasoro minimal model. The $\mathcal{S}$-matrix for the minimal model is given by (see, e.g., (18.91) of \cite{DiFrancesco:1997nk})
\begin{align}
\mathcal{S}_{(l;p)}^{~~ (l';p')} = \sqrt{\frac{2}{(k+2)(k+3)}}\, \sin \left( \frac{\pi (2 l + 1) (2 l' +1)}{k+2}\right)\, \sin \left( \frac{\pi (2 p + 1) (2 p' +1)}{k+3}\right)\ ,  \label{minimalS}
\end{align}
up to a sign factor $(-1)^{(2 l + 2 p)(2 l ' + 2 p ')} $.
The vacuum partition function calculated from the coset CFT is
\begin{align}
Z_\text{CFT} = \left|\mathcal{S}_{(0;0)}^{~~ (0;0)} \right|^2 \, .
\end{align}
As explained before, the leading order in $1/c^{(g)}$ comes from the $\mathcal{S}$-matrix of SU$(2)$ at level $k$. However, the term proportional to $\log c^{(g)}$ arises also from the $\mathcal{S}$-matrix of SU$(2)$ at level $k+1$ as
\begin{align}
\sin \left( \frac{\pi}{k+3}\right) \simeq \sin \left( \frac{\pi}{1 + \i \,6 \,  /c^{(g)}}\right) \simeq
\sin \left( \pi -  \i \, \frac{6 \,  \pi  }{c^{(g)}}\right) \simeq \i \, \frac{6 \,    \pi}{c^{(g)}}\ .
\end{align}
Thus, the total contribution can be read off as
\begin{align}
I_\text{CFT}^{(1)} = 1 \, .
\end{align}
It still does not match with \eqref{IG1}, but it is natural as we do not include perturbative matter in the gravity side.

We can do a similar analysis for generic $N$. To examine the effects at the next leading order in $1/c^{(g)}$, we consider the coset CFT \eqref{cosetc} as a concrete example. The modular $\mathcal{S}$-matrix is written as
\begin{align}
\mathcal{S}_{(\mu;\nu)}^{~~(\mu ';\nu ')} = \mathcal{S}^{(k) \mu ' }_{\mu } \mathcal{S}^{(1) \xi '}_{\xi } \mathcal{S}^{(k+1) \nu ' }_{\nu } \, , \qquad
\mathcal{S}^{(k)\mu '}_\mu=K\sum_{w\in W}\ep(w)\,e^{-\frac{2\pi \i}{k+N}(w(\mu+\rho),\mu '+\rho)} \, , \label{wms}
\end{align}
see \eqref{SN} and also \cite{DiFrancesco:1997nk} for more details. 
As explained before, $\xi$ is uniquely fixed by $\mu,\nu$ through the selection rule of the coset, and the same is true for $\xi '$.
Thus the vacuum partition function computed from the coset CFT \eqref{cosetc} is
\begin{align}
Z_\text{CFT} = \left|\mathcal{S}_{(0;0)}^{~~ (0;0)} \right|^2 = \left|\mathcal{S}^{(k) 0 }_{0}\, \mathcal{S}^{(1) 0}_{0}\, \mathcal{S}^{(k+1) 0 }_{0}\right|^2 \, , \label{ZCFT}
\end{align}
where
\begin{align}
\mathcal{S}^{(k) 0}_{0} =  \frac{1}{\sqrt{N}}\frac{1}{(k +N)^{\frac{N-1}{2}}} \prod_{p=1}^{N-1} \left[ 2 \sin\left( \frac{\pi p}{k+N}\right)\right]^{N-p} \, .
\end{align}
The leading order expression was already found to reproduce that from the higher-spin computation in section \ref{sec:hs}. The contribution proportional to $\log c^{(g)}$ is computed as
\begin{align}
I_\text{CFT}^{(1)} = 
- (N-1) + 2 \sum_{p=1}^{N-1} (N-p) = N^2  - 2 N + 1 \, .
\end{align}
Here we remark that the vacuum partition function computed from the higher-spin gravity at the same order is
\begin{align}
I_\text{G}^{(1)} = N^2 -1 \, .
\end{align}
This result can be obtained from \cite{Anninos:2020hfj} or $1/k$ expansion of \eqref{SN}. The two computations does not match with each other as in the case of $N=2$, which suggests that perturbative matters should be included.

Up to now, we have considered only contributions from (higher-spin) gravity. 
Here we would like to include the effects of additional matters. Again we consider the analytic continuation of the Gaberdiel-Gopakumar duality discussed in section \ref{sec:GG}. In this case, the additional matter is a complex scalar with dual conformal dimension $\Delta = 1 - \lambda = 1 - N$. The one-loop correction of the partition function of a real scalar field (i.e. $s=0$) with mass $m^2$ can be read off from e.g., (1.12) of \cite{Anninos:2020hfj}. Now $ L^2 m^2  = 1 - \lambda^2$. For generic $m^2$, there is no term proportional to $\log c^{(g)}$, However, now $\lambda$ is an integer number, and the one-loop partition function diverges for the case. There is a similar divergence also for massless higher-spin fields and a careful treatment of regularization leads to the term proportional to $\log c^{(g)}$. We expect that a term proportional to $\log c^{(g)}$ arises also for a massive scalar field with integer $\lambda$, but currently we do not know any prescription to regularize the divergence.
Therefore, we cannot conclude whether the partition function at the order of $\log c^{(g)}$ agrees or not between the CFT and gravity computations even for the case of analytic continuation of the Gaberdiel-Gopakumar duality.%
\footnote{Instead of working on the limit \eqref{limit0} of the coset model \eqref{cosetc}, we may consider the 't Hooft limit explained in the previous section.  However, as argued in appendix H.3 of  \cite{Anninos:2020hfj}, there are different difficulties in the computations at the one-loop level. It is a future work to resolve these issues with the partition function at the higher order in $1/c^{(g)}$ expansion.}

\section{Conclusion and discussions}
\label{sec:conclusion}

In this paper, we proposed a new example of dS/CFT correspondence for three-dimensional de Sitter spaces. Our duality agrees with the general expectation of dS/CFT such that the classical gravity limit corresponds to the infinitely large imaginary valued central charge. The proposed duality relation is labeled by an integer $N \, (\geq 2)$ and is summarized as  (\ref{dualitytwo}) for $N=2$ and (\ref{dualityN}) for $N>2$. The gravity theory is given by a higher-spin theory, which includes gauge fields of spin $s = 2,3,\ldots,N$. In particular, the special case $N=2$ corresponds to the Einstein gravity.  

We argued that the dual CFTs are in a class of CFTs which include the large central charge limit $k\to -N$  of the SU$(N)_k$ WZW model given by 
(\ref{centtwo}) and (\ref{centNa}). This part gives the dominant contributions to physical quantities due to the large central charge limit and is dual to the gravity degrees of freedom. Indeed, we showed explicitly that the semiclassical partition functions of Einstein gravity and higher-spin gravity with various excitations on $\mathbb{S}^3$ perfectly agree with those computed from the dual CFTs. In the higher-spin gravity, such solutions with excitations are realized by introducing conical defects in $\mathbb{S}^3$ and we presented explicit solutions in the Chern-Simons formulation of higher-spin gravity. On the other hand, their CFT counterparts are products of the modular $\mathcal{S}$-matrices, which exponentially enhance to reproduce the classical (higher-spin) gravity result in the $k\to -N$ limit. We also pointed out that our 
dS/CFT correspondence may also be interpreted as an analytical continuation of the higher-spin holography (Gaberdiel-Gopakumar duality). Interestingly, the triality relation known for this duality predicts that the gravity dual of our two-dimensional CFT has a $W_N$ asymptotic symmetry, which gives a further evidence that it is a classical higher-spin gravity with spin $s=2,3,\ldots,N$.

Moreover, we explored this new dS/CFT duality further in two different directions. As one of them, we studied partition functions, two-point functions, and holographic entanglement entropy to promote the above duality in the Euclidean version of dS/CFT to the Lorentzian one. In dS/CFT, we identify the future infinity of dS$_3$ with the manifold $\mathbb{S}^2$ where the dual CFT lives. We computed partition function of Liouville/Toda theory on the $\mathbb{S}^2$ and found that the result is consistent with the above picture.
Moreover, we showed that two-point functions in the dual CFT on $\mathbb{S}^2$ agree with those obtained from the dS$_3$ under the geodesic approximation, with an exotic rule in the CFT that the UV cutoff is taken to be an imaginary value. This prescription peculiar to the dS/CFT is also supported from the Hartle-Hawking prescription which creates the de Sitter space from nothing via the tunneling effect. This result of two-point functions implies that we can obtain the Lorentzian version of dS/CFT by an appropriate analytical continuation from the Euclidean version, where the cutoff parameter plays the role of an emergent time. This is a quite intriguing setup of dS/CFT because the Lorentzian dynamics arises from the topological theory of Chern-Simons theory on $\mathbb{S}^3$. In a similar way, we can calculate the entanglement entropy for the dual CFT on $\mathbb{S}^2$. This can again be perfectly reproduced from the gravity calculation, namely, the holographic entanglement entropy. We furthermore confirmed this result in the light of codimension-two and brane-world holography for a four-dimensional AdS wedge spacetime. It will be a very interesting future problem to study this continuation from the Euclidean space to Lorentzian dS further to fully understand the mechanism of emergent time in dS/CFT.

Another direction we explored is the quantum corrections. In the first half of this paper, we confirmed the perfect matching between the classical (higher-spin) gravity partition functions and those derived from dual CFTs.
As a next step, we naturally ask whether we can have a similar matching for one-loop quantum gravity corrections. This is a much more nontrivial question because we need to know the full matter field contents in the gravity theory, which are expected to be affected by the details of two-dimensional CFTs we choose, i.e., the choices of $M_\text{CFT}$ in (\ref{dualitytwo}) and (\ref{dualityN}). In this paper, we made a first step toward this problem. We evaluated the one-loop corrections of the gravity partition function as CFT predictions. We found that both the pure SU$(N)_k$ and the $W_N$-minimal model $W_{N,k}$ \eqref{cosetc} produce different one-loop corrections. We also noted that neither of them does not agree with the pure (higher-spin) gravity contributions. This implies a presence of additional matter fields, which produce extra contributions to the one-loop corrections. We would like to leave further studies for a future problem.

\subsection*{Acknowledgements}

We are grateful to Heng-Yu Chen, Shiraz Minwalla, K. Narayan, Andrew Strominger, Kenta Suzuki and Sandip Trivedi for useful comments and conversations.
This work was supported by JSPS Grant-in-Aid for Scientific Research (A) No.\,21H04469, and
Grant-in-Aid for Transformative Research Areas (A) ``Extreme Universe''
No.\,21H05182, No.\,21H05187 and No.\,21H05190.
The work of Y.\,H. was supported in part by the JSPS Grant-in-Aid for Scientific Research (B) No.19H01896. The work of T.\,N. was supported in part by the JSPS Grant-in-Aid for Scientific Research (C) No.19K03863.
T.\,T. is supported by the Simons Foundation through the ``It from Qubit'' collaboration, Inamori Research Institute for Science and 
World Premier International Research Center Initiative (WPI Initiative) from the Japan Ministry of Education, Culture, Sports, Science and Technology (MEXT). 
We would also like to thank the online conferences: ``Indian Strings Meeting 2021'' hosted by IIT Roorkee, and the YITP workshop:
``KIAS-YITP 2021: String Theory and Quantum Gravity'' (YITP-W-21-18), hosted by YITP, Kyoto U, for stimulating discussions and comments from participants, where this work was presented.

\appendix

\section{Description by Liouville/Toda CFTs}
\label{app:DS}

As mentioned in the introduction, we can use Liouville/Toda CFT with $b \to 0$ instead of SU(N) WZW model with $k \to -N$ since the both provide equivalent descriptions at the leading order in $1/c^{(g)}$.
In this appendix, we show that the results obtained in the main context can be reproduced also from the Liouville/Toda CFT in the limit. The analysis provides different useful viewpoints of our dS$_3/$CFT$_2$ duality.
Moreover, one may feel that the critical level limit of  SU$(N)$ WZW model is not rigorously justified.
Physical quantities, such as, $\mathcal{S}$-matrix in \eqref{SU2S} (or \eqref{SN}) are obtained for integer $k$ and analytic continuations are taken with respect to $k$. 
Another aim of this appendix is to give evidence for our expression at the critical level limit.

\subsection{Liouville CFT}

The action of Liouville CFT is given by
\begin{align}
I = \frac{1}{2 \pi} \int \d^2 w \sqrt{g}\left( \bar \partial \phi\, \partial \phi + \frac{Q}{4}\, \mathcal{R}\, \phi + \mu\, e^{2 b \phi} \right)
\label{Liouvillea}
\end{align}
with $g_{ab}$ as the metric of world-sheet, $g$ as $\sqrt{\det g_{ab}}$, and $\mathcal{R}$ as the Ricci curvature with respect to $g_{ab}$.
The background charge $Q$ is related to the central charge $c$ and the parameter $b$ as
\begin{align}\label{eq:liouvillec}
    c = 1 + 6 Q^2 \, , \quad Q = b + \frac{1}{b} \, . 
\end{align}
The Liouville CFT is known to have the self-duality under $b \to 1/b$ \cite{Zamolodchikov:1995aa}. In fact, we can see that the central charge is invariant under the self-duality.
Since the Liouville CFT is equivalent to the coset \eqref{cosetc}, the relation between the parameter should be given by
\begin{align}
c=1+6Q^2=1-\frac{6}{(k+2)(k+3)} \, .
\end{align}
In particular, the $k \to -2$ limit of the coset \eqref{cosetc} with $N=2$ corresponds to $b \to 0$ (or $b \to \infty$) limit of Liouville CFT.

The vertex operators in the Liouville CFT are of the form $V_\alpha = \exp (2 \alpha \phi)$ and the conformal weights are $h_\alpha = \alpha (Q -\alpha)$.
Physical states correspond to operators with
\begin{align}
\alpha = \frac{Q}{2} + \i \, p \quad (p \in \mathbb{R}) \,  , \label{alphap}
\end{align}
while degenerate operators have specific valued parameters 
\begin{align}
\alpha_{r,s} = \frac{b (1-r) + b^{-1}(1 -s) }{2} \label{alphars}
\end{align}
with $r,s =1,2,\ldots$.
In particular, the identity operator corresponds to $(r,s) = (1,1)$. The $\mathcal{S}$-matrix between the degenerate and physical ones is given by \cite{Zamolodchikov:2001ah}
\begin{align}
\mathcal{S}_{(r,s)}^{~~p} \propto \sinh(2\pi \, r\, p\, b)\,\sinh(2\pi \, s\,p/b) \, . \label{LiouvilleS}
\end{align}
On the other hand, the Liouville CFT is equivalent to the coset CFT \eqref{cosetc} with $N=2$, and the modular $\mathcal{S}$-matrix for the coset is given by \eqref{minimalS}.
The $\mathcal{S}$-matrix between the primary states $(l;l')$ and $(j;j)$ reads
\begin{align}
\begin{aligned}
 \mathcal{S}_{(l;l')}^{~~(j;j)} &\propto \sin\left(\frac{\pi}{k+2}(2 l+1)(2j+1)\right)\,
\sin\left(\frac{\pi}{k+3}(2 l' +1)(2j+1)\right)\\
&=\sin\left(\pi\, Q\,b\, (2 l+1)(2j+1)\right)\,\sin\left(\frac{\pi\, Q}{b}\,(2 l'+1)(2j+1)\right)\\
&=\sinh(2\pi\, r\, p\, b)\sinh(2\pi\, s\, p/b) \,.
\end{aligned}
\end{align}
Here we relate the Liouville momentum $p$ and SU$(2)$ quantum number $j$ as 
\ba
h=\frac{Q^2}{4}+p^2=\frac{j(j+1)}{k+2}-\frac{j(j+1)}{k+3} \,,
\ea
which leads to 
\ba
p=\frac{\i\, Q}{2}\,(2j+1) \, .
\ea
Similarly, we set
\begin{align}
r= 2 l + 1 \, , \quad s= 2 l' +1 \, .\label{rs}
\end{align}
In this way, we reproduce \eqref{LiouvilleS} from the  $\mathcal{S}$-matrix for the coset \eqref{cosetc}.

Even though we already explained the equivalence between the coset CFT \eqref{cosetc} with $N=2$ and the Liouville CFT, it is instructive to confirm that the expected gravity partition function can be derived from the Liouville $\mathcal{S}$-matrix correctly.  In the Liouville theory,
the $\mathcal{S}$-matrix is given by \eqref{LiouvilleS}.
We are interested in the matrix element corresponding to $\mathcal{S}_{l '}^{~j}$ for SU$(2)$ WZW model and thus we set $p \to \frac{\i\,Q}{2} (2 j +1) \simeq \frac{\i}{2b} (2 j +1)$
in the large $c$ limit, i.e., $b\to 0$.
To continue from the normal Liouville CFT to that for de Sitter CFT, 
the value of $b$ gets complex as 
\ba
\i \, c^{(g)}\simeq \frac{6}{b^2} \, . \label{cdsb}
\ea
This leads to
\begin{align}
    \begin{aligned}
        \left|\mathcal{S}_{(r,s)}^{~~\frac{\i\,Q}{2} (2 j+1)}\right|^2
            &\sim 
            \exp\left[-\frac{2\pi \i\, (2 l' +1) (2 j+1)}{b^2}\right]\\ &\sim \exp\left[\frac{\pi}{3}c^{(g)}\sqrt{1 - 8 G_N E_{l'}}\sqrt{1 - 8 G_N E_{j}}\right] \, ,
    \end{aligned}
\end{align}
where we have used \eqref{rs} and \eqref{E2j}.
The above expression reproduces the previous result (\ref{twowa}) from the classical gravity on $\mathbb{S}^3$.

\subsection{Toda CFT}

In this subsection, we extend the previous analysis to the case with SU$(N)$ at the critical level limit \eqref{dualityN}.
The coset CFT \eqref{cosetc} was shown to have $W_N$-symmetry with spin-$s$ currents $(s=2,3,\ldots,N)$ \cite{Arakawa:2018iyk}.
On the other hand, a definition of $W_N$-algebra is given by so-called Drinfeld-Sokolov (DS) reduction of the SU$(N)$ current algebra at level $k_\text{DS}$, see, e.g., \cite{Bouwknegt:1992wg}.
The level $k$ of the latter coset CFT is related to $k_\text{DS}$ as
\ba
\frac{1}{k+N}=\frac{1}{k_\text{DS}+N}-1 \, .
\ea
In terms of the DS reduction, the triality relations (\ref{traa}) and 
(\ref{trab}) are rewritten as
\begin{align}
 (a)\quad k'_\text{DS}+N'&=\frac{1}{k_\text{DS}+N} \, , & N'&=N \, ,\label{DSa}\\
 (b)\quad  k'_\text{DS}+N'&=1-k_\text{DS}-N\, ,& N'&=\frac{N}{N+k_\text{DS}}-N \, .
\label{DSb}
\end{align}
The relation $(a)$ for the $W_N$-algebra is known as Feigin-Frenkel duality \cite{FF}.

Another realization of $W_N$-algebra is via free field realization, which implies that the symmetry algebra of Toda CFT is the $W_N$-algebra.
The action of Toda CFT is given by%
\footnote{Note that, in the standard notation for Toda CFT as expressed by the action, setting $N=2$ does not simply reduce to the Liouville CFT with the action \eqref{Liouvillea} and some change of parameters is required.}
\begin{align}
    I = \frac{1}{2 \pi} \int \d^2 w \sqrt{g}\left( \frac{G^{ij}}{2}\, \bar \partial \phi_i\, \partial \phi_j + \frac{ \mathcal{R} }{4} (b + b^{-1})\sum_{i=1}^{N-1}\phi^i + \mu \sum_{i=1}^{N-1} e^{b \phi_i} \right) \, ,
\end{align}
where $G_{ij}$ is the Cartan matrix of SU$(N)$ and the inverse is defined via $G^{ij}G_{jl} = \delta ^i_{~l}$. The index for the Toda fields $\phi_j$ is raised as $\phi^i = G^{ij}\phi_j$.
The central charge can be expressed as 
\begin{align}\label{Todac}
c = N -1 + 12 (Q,Q) \, , \qquad Q = \left(b + \frac{1}{b} \right)\rho \, , \qquad b^2 = - \frac{1}{k_\text{DS} + N} \, ,
\end{align}
where $\rho$ is the Weyl vector defined in \eqref{rho}.
The Feigin-Frenkel duality \eqref{DSa} is realized as the self-duality under $b \to 1/b$ in the Toda field theory \cite{Fateev:2007ab}.

The vertex operators of the Toda CFT can be given by $V_\alpha = \exp (\sum_{i=1}^{N}\alpha_i \varphi_i)$ with $\phi_j = \varphi_j - \varphi_{j+1}$. The conformal dimension of the operator is $h_\alpha = (\alpha , 2 Q - \alpha)/2$ with $\alpha = \sum_{j=1}^N \alpha_j e_j$.
For Liouville CFT, we have used two types of spectra with $\alpha$ in \eqref{alphap} and \eqref{alphars}. In the Toda CFT, there are $N-1$ momenta, and we can use one of the two choices for each momentum. For our purpose, we only use
\begin{align}
\alpha = Q + \i \, p \quad (p_j \in \mathbb{R}) 
\end{align}
for all momenta or $\alpha = \alpha_{\mu,\nu}$ with
\begin{align}
\alpha_{\mu,\nu} = Q - b(\mu + \rho) - b^{-1}(\nu + \rho) \, . 
\end{align}
Here Young diagrams are represented by $\mu$ and $\nu$. In particular, $\mu = \nu = 0$ corresponds to the identity operator.
The modular $\mathcal{S}$-matrix between the two types of momenta was obtained in \cite{Drukker:2010jp} as
\begin{align}
 \mathcal{S}_{(\mu,\nu)}^{~~p} \propto \sum_{w \in W}  \epsilon (w) e^{- 2 \pi b (w (\mu + \rho) , p )}\sum_{w ' \in W}  \epsilon (w ') e^{- \frac{2 \pi}{b}  (w ' (\nu + \rho) , p )} \, , 
\end{align}
where the sums are taken over the elements of SU$(N)$ Weyl group and $\epsilon (w) = \pm 1$.
As in the Liouville case, we can reproduce the expression from 
$S_{(\mu;\nu)}^{~~(\sigma;\sigma)}$ in \eqref{wms} with $p = \i \, (b + b^{-1}) (\sigma + \rho)$. In order to reproduce $\mathcal{S}_{\nu}^{~\sigma}$ for SU$(N)$ WZW model near the critical level, we take $p \to \i \, (b + b^{-1}) (\sigma + \rho) \simeq \i \, (\sigma + \rho) /b $ with $b \to 0$. Setting $b$ as
\ba
\i \, c^{(g)}\simeq \frac{12(\rho , \rho )}{b^2} \label{cds}
\ea
as in \eqref{cdsb}, we find
\ba
\left|\mathcal{S}_{(\mu,\nu)}^{~~\i\, (b + b^{-1})( \sigma + \rho)}\right|^2\sim e^{-\frac{4\pi \, \i \, ( \nu + \rho , \sigma + \rho)}{b^2}}\sim e^{\frac{\pi}{3}c^{(g)}\frac{(\nu + \rho ,\sigma + \rho)}{(\rho ,\rho)} } \, , \label{Todar}
\ea
which reproduces the previous result (\ref{cft}).

\section{Geodesic from Wilson line} \label{app:Wilsonline}
Here we work out the relation between the expectation value of a Wilson line and the geodesic distance, by extending the argument in AdS$_3/$CFT$_2$ \cite{Ammon:2013hba} to our dS$_3/$CFT$_2$.
This also provides a prescription for computing the holographic entanglement entropy in dS$_3/$CFT$_2$.

\subsection{Wilson line calculation}
Consider an SU$(2)$-valued field $U$. Wilson line along a curve $\gamma(s),\,s\in [s_i,s_f]$ with two boundary conditions $\ket{U_i},\ket{U_f}$ can be expressed as in the AdS case of \cite{Ammon:2013hba}
\begin{align}\begin{aligned}\label{eq:Wilson}
  \bra{U_f}\mathcal{P}\exp\left(\int_{\gamma}A\right)\mathcal{P}\exp\left(\int_{\gamma}\bar{A}\right)\ket{U_i} 
  =\int\mathcal{D} U\,\mathcal{D} P\,\mathcal{D} \lambda\,\exp\left(-I_{\gamma}[U,P,\lambda;A,\bar{A}]\right) \, ,
\end{aligned}\end{align}
where
\begin{align}\begin{aligned}\label{eq:mpaction}
  I_{\gamma} [U,P,\lambda;A,\bar{A}] =\int_{s_i}^{s_f} \d s\left[\Tr(PU^{-1}D_sU)+\lambda(s)\left(\Tr P^2-c_2\right)\right] \, ,
\end{aligned}\end{align}
and $\mathfrak{su}(2)$-valued field $P$ is a canonical momentum conjugate to $U$. The covariant derivative $D_s$ defined by 
\begin{align}
  D_sU=\frac{\d U}{\d s}+A_sU-U\bar{A}_s \, ,\qquad A_s\equiv A_\mu\frac{\d x^{\mu}}{\d s},\ \qquad \bar{A}_s\equiv A_\mu\frac{\d x^\mu}{\d s} \, .
\end{align}
Here $\lambda(s)$ plays the role of Lagrange multiplier constraining $\Tr P^2$ to the quadratic Casimir $c_2$.\footnote{Note that $c_2$ is negative for $\SU(2)$ since $P$ is anti-Hermitian.}
The action \eqref{eq:mpaction} is invariant under a local gauge transformation
\begin{align}
  U(s)&\to L\left(\gamma(s)\right)\,U(s)\,R\left(\gamma(s)\right) \, , \\
  P(s)&\to R^{-1}\left(\gamma(s)\right)\,P(s)\,R\left(\gamma(s)\right) \, .
\end{align}
In the AdS$_3$ case, it is shown in \cite{Ammon:2013hba} that on-shell the action $S_\gamma$ (divided by $\sqrt{2|c_2|}$) with boundary conditions $U(s_i)=\mathbbm{1}$ and $U(s_f)=\mathbbm{1}$ reduces to the length of the geodesic connecting two edge points $\gamma(s_i)$ and $\gamma(s_f)$. Below we will confirm that the same is true for our dS$_3$.

The equations of motion for \eqref{eq:mpaction} are 
\begin{align}\label{eq:mpEOM}
  U^{-1}D_sU+2\lambda P=0 \, ,\qquad \frac{\d P}{\d s}+[\bar{A}_s,P]=0 \, ,\qquad \Tr P^2=c_2 \, ,
\end{align}
and the on-shell action becomes 
\begin{align}\label{eq:mponshell}
  I_{\text{on-shell}}=-2c_2\int_{s_i}^{s_f} \d s\, \lambda(s)=-2c_2\,\Delta\alpha \, ,
\end{align}
where $\Delta\alpha=\alpha(s_f)-\alpha(s_i)$ and $\alpha$ satisfies $\d\alpha/\d s=\lambda(s)$. Now let $A$ and $\bar{A}$ take the forms
\begin{align}
  A=L\,\d L^{-1}\, ,\qquad \bar{A}=R^{-1}\,\d R \, ,\qquad (L,R)\in \text{SU}(2)\times \text{SU}(2) \, .
\end{align}
We can obtain this solution by performing a gauge transformation from a trivial solution $A=\bar{A}=0$, so we first solve \eqref{eq:mpEOM} in this case, giving 
\begin{align}
  P(s)=P_0 \, ,\qquad U(s)=U_0(s)\equiv u_0\exp\left(-2\alpha(s)P_0\right)\, ,
\end{align}
where $P_0$ and $u_0$ are constant operators. We perform a gauge transformation $(L,R)$, then we have 
\begin{align}
  U(s)&=L\left(\gamma(s)\right)\,U_0(s)\,R\left(\gamma(s)\right)\, , \\ P(s)&=R^{-1}\left(\gamma(s)\right)\,P_0\,R\left(\gamma(s)\right) \, . 
\end{align}
We impose boundary conditions for $U$ as $U(s_i)=U_i$ and $U(s_f)=U_f$, then we have 
\begin{align}
  U_i&=L\left(\gamma(s_i)\right)\,u_0\,e^{-2\alpha(s_i)P_0}\,R\left(\gamma(s_i)\right)\, , \\
  U_f&=L\left(\gamma(s_f)\right)\,u_0\,e^{-2\alpha(s_f)P_0}\,R\left(\gamma(s_f)\right)\, .
\end{align}
This leads to 
\begin{align}
  e^{2\Delta \alpha P_0}=R\left(\gamma(s_f)\right)\,U^{-1}_f\,L\left(\gamma(s_f)\right)\left[R\left(\gamma(s_i)\right)\,U_i^{-1}\,L\left(\gamma(s_i)\right)\right]^{-1}\, .
\end{align}
To evaluate the value of $\Delta\alpha$, we take trace for some representation of SU$(2)$. We adopt here the fundamental representation, then 
\begin{align}
  \tr_{\text{f}}\,e^{2\Delta \alpha P_0}=2\cos(\Delta \alpha\sqrt{2|c_2|}) \ ,
\end{align}
because $P_0$ has eigenvalues $\pm \i \, \sqrt{|c_2|/2}$. Solving this equation for $\Delta \alpha$ and substituting it to \eqref{eq:mponshell} we obtain
\begin{align}\begin{aligned}\label{eq:mpactionbc}
  I_{\text{on-shell}}=\sqrt{2|c_2|}\,\cos^{-1}\left[\frac{1}{2}\right.\tr_{\text{f}}\left(R\left(\gamma(s_f)\right)\,U^{-1}_f\,L\left(\gamma(s_f)\right)\right.
\left[R\left(\gamma(s_i)\right)\,U_i^{-1}\,L\left(\gamma(s_i)\right)\right]^{-1}\Bigr)\Bigr] \, .
\end{aligned}\end{align}
Therefore we can obtain the value if we specify the connections $L\left(\gamma(s)\right),R\left(\gamma(s)\right)$, the end points of the Wilson line $\gamma(s_i),\gamma(s_f)$ and the boundary conditions.

\subsection{Euclidean \texorpdfstring{$\text{dS}_3$}{dS3} with boundary: \texorpdfstring{$\BB^3$}{B3}}
Here we calculate a Wilson line in $\BB^3$ that ends on the boundary $\mathbb{S}^2$. First, we take the global patch,
\begin{align}\label{eq:globalp}
    \d s^2=\d r^2+\sin^2 r\,\left(\d\theta^2+\sin^2 \theta\, \d\phi^2\right) \, ,
\end{align}
of $\mathbb{S}^3$ and cut along an equator $r=\pi/2$ to obtain $\BB^3$. Note that the radius of $\mathbb{S}^3$ is fixed to $1$.

It is easier to use the static patch,
\begin{align}\label{eq:staticp}
    \d s^2=\cos^2\rho \,\d\tau^2+\d\rho^2+\sin^2\rho \,\d\phi^2 \, ,
\end{align}
in order to calculate the Wilson line. Note that $0\le\rho\le\pi/2$. The associated gauge connections with \eqref{eq:staticp} are \cite{Castro:2020smu}
\begin{align}
  A&=L\,\d L^{-1} \, ,\qquad L=e^{-\i\, \rho L_2}\,e^{-\i\, (\phi+\tau)L_3}\, , \\
  \bar{A}&=R^{-1}\,\d R \, ,\qquad R=e^{\i\, (\phi-\tau)L_3}\,e^{-\i \,\rho L_2}\, .
\end{align}
The on-shell action \eqref{eq:mpactionbc} with the endpoints 
\begin{align*}
    &\rho(s_i)=\rho_i,\qquad\rho(s_f)=\rho_f \, ,\\
    &\tau(s_f)-\tau(s_i)=\Delta \tau \, ,\\
    &\phi(s_f)-\phi(s_i)=\Delta\phi \, ,
\end{align*}
becomes 
\begin{align}\begin{aligned}
    I_{\text{on-shell}}=\sqrt{2|c_2|}\,\cos^{-1}\left(\cos\Delta\tau\,\cos\rho_i\,\cos\rho_f +\cos\Delta\phi\,\sin\rho_i\,\sin\rho_f\right) \, ,
\end{aligned}\end{align}
which is indeed identical to the geodesic distance between $\gamma(s_i)$ and $\gamma(s_f)$ in the $\BB^3$.

The relations between the global patch \eqref{eq:globalp} and the static patch \eqref{eq:staticp} are given by 
\begin{align}
    \sin\rho=\sin r\sin\theta \, ,\qquad \cos^2\tau=\frac{\cos^2 r}{1-\sin^2r\sin^2\theta}\,.
\end{align}
On the boundary $r=\pi/2$, this reduces to 
\begin{align}
    \sin\rho=\sin\theta \, ,\qquad \cos\tau=0 \, .
\end{align}
Therefore, the endpoints of a Wilson line on the boundary satisfy
\begin{align}
    \tau=\frac{\pi}{2}\, ,\ \rho=\theta \qquad \text{or}\qquad \tau=\frac{3\pi}{2}\, ,\ \rho=\pi-\theta\ .
\end{align}
Thus, the geodesic distance becomes 
\begin{align}\label{eq:cccss}
    \frac{I_{\text{on-shell}}}{\sqrt{2|c_2|}}=\cos^{-1}\left(\cos\theta_i\,\cos\theta_f + \cos\Delta\phi\,\sin\theta_i\,\sin\theta_f\right) \ ,
\end{align}
in the global patch.
Let $\bm{n}_i,\bm{n}_f$ be vectors representing the points on $\mathbb{S}^2$:
\begin{align}
    \bm{n}_i&=(\sin\theta_i\cos\phi_i,\ \sin\theta_i\sin\phi_i,\ \cos\theta_i)^{\mathrm{T}} \, , \\
    \bm{n}_f&=(\sin\theta_f\cos\phi_f,\ \sin\theta_f\sin\phi_f,\ \cos\theta_f)^{\mathrm{T}} \, ,
\end{align}
then \eqref{eq:cccss} reduces to 
\begin{align}
    \frac{I_{\text{on-shell}}}{\sqrt{2|c_2|}}=\cos^{-1}(\bm n_i\cdot\bm n_f) \, ,
\end{align}
which represents the length of the arc between $\gamma(s_i)$ and $\gamma(s_f)$ in the $\BB^3$.

\subsection{Lorentzian dS\texorpdfstring{$_3$}{3}: Poincar{\'e} coordinates}

Next, we consider Lorentzian dS$_3$ spacetime in the Poincar{\'e} patch:
\begin{align}\label{eq:dspoi}
  \d s^2=\frac{-\d z^2+\d w\, \d\bar{w}}{z^2} \, ,
\end{align}
where $w=x+\i \, y$. Three-dimensional dS gravity can be formulated by $\text{SO}(3,1)\simeq \text{SL}(2,\C)$ Chern-Simons theory. The associated gauge connections to the metric \eqref{eq:dspoi} are
\begin{align}
  A&=L\,\d L^{-1} \, ,\qquad L=e^{-\frac{\i \, w}{z}J_1}\,e^{\log z\cdot J_0} \, , \\
  \bar{A}&=R^{-1}\,\d R \, ,\quad R=e^{\log z\cdot J_0}\,e^{\frac{\i\, \bar{w}}{z}J_{-1}} \, ,
\end{align}
where $J_a$ are the generators of $\mathfrak{sl}(2)$. We fix the coordinates of the end points as 
\begin{align}
  z(s_i)=z_i \, ,\quad z(s_f)=z_f \, ,\quad w(s_i)=w_i \, ,\quad w(s_f)=w_f \, ,
\end{align}
and impose the boundary conditions 
\begin{align}
  U_i=U_f=\mathbbm{1} \, ,
\end{align}
then 
\begin{align}\begin{aligned}\label{eq:actionPoincare}
  I_{\text{on-shell}}=\sqrt{2c_2}\cos^{-1}\left(\frac{z_i^2+z_f^2-|\Delta w|^2}{2z_i\,z_f}\right)\, .
\end{aligned}\end{align}
This form is identical to the geodesic distance between $\gamma(s_i)$ and $\gamma(s_f)$ in Poincar{\'e} dS$_3$.

If the Wilson line ends on the future asymptotic boundary, by fixing  $z_i=z_f\equiv\epsilon$,
\begin{align}\begin{aligned}
  I_{\text{on-shell}}&=\sqrt{2c_2}\cos^{-1}\left(1-\frac{|\Delta w|^2}{2\epsilon^2}\right).
\end{aligned}\end{align}
Therefore the on-shell action is real only if $|\Delta w|^2/2\epsilon^2<1$.
In particular, when $|\Delta w|^2/2\epsilon^2\ll1$, 
\begin{align} \label{eq:geodesicpoincare}
   I_{\text{on-shell}}\simeq\sqrt{\frac{c_2}{2}}\cdot\frac{|\Delta w|^2}{\epsilon^2}\,.
\end{align}
On the other hand, if $|\Delta w|^2/2\epsilon^2\gg1$, then 
\begin{align}
    \frac{I_{\text{on-shell}}}{\sqrt{2c_2}}\simeq \pi\pm \i \, \log\left(\frac{|\Delta w|^2}{\epsilon^2}\right) \, .
\end{align}

\subsection{Lorentzian dS\texorpdfstring{$_3$}{3}: global coordinates}

Let us finally study the global patch,
\begin{align}
    \d s^2=-\d T^2+\cosh^2T\left(\d\psi^2+\sin^2\psi\, \d\phi^2\right)\, ,
\end{align}
which is related to the Poincar\'{e} patch via the coordinate transformation:
\begin{align}
    \frac{1}{z}=\sinh T+\cosh T\cos\psi,\qquad |w|=\frac{\sin\psi}{\tanh T+\cos\psi}\, .
\end{align}
Fixing the end points as
\begin{align}
    T(s_i)&=T_i\, , & \psi(s_i)&=\psi_i\, , & \phi(s_i)&=\phi_0\, , \\
    T(s_f)&=T_f\, , & \psi(s_f)&=\psi_f\, , & \phi(s_f)&=\phi_0 \, , 
\end{align}
the on-shell action
\eqref{eq:geodesicpoincare} becomes in the limits $T_i,T_f\gg1$
\begin{align}\label{eq:geodesicstatic}
    \frac{I_{\text{on-shell}}}{\sqrt{2c_2}}\simeq \pi\pm \i \, \log\left[e^{T_i+T_f}\sin^2\left(\frac{\psi_f-\psi_i}{2}\right)\right]\, ,
\end{align}
which agrees with the result of the ordinary calculation of the geodesic \eqref{eq:HHgeodesic}.


\providecommand{\href}[2]{#2}\begingroup\raggedright\endgroup

\end{document}